\documentclass[12pt]{article}

\usepackage[T1]{fontenc}
\usepackage[margin=1.2in]{geometry}
\usepackage{amsmath,amssymb,amsthm,mathtools}
\usepackage{booktabs,array}
\usepackage{enumitem}
\usepackage{microtype}
\usepackage{setspace}
\usepackage{natbib}
\usepackage{titlesec}
\usepackage{tikz}
\usetikzlibrary{arrows.meta,positioning,fit}
\usepackage{hyperref}
\hypersetup{colorlinks=true, linkcolor=blue, citecolor=blue, urlcolor=blue}

\setlist{nosep}
\titleformat{\section}{\centering\normalfont\normalsize\scshape}{\thesection.}{0.5em}{}
\titleformat{\subsection}{\centering\normalfont\normalsize}{\thesubsection}{0.6em}{\itshape}
\titleformat{\paragraph}[runin]{\normalfont\normalsize\itshape}{}{0pt}{}
\titlespacing*{\section}{0pt}{3.4ex plus 1ex minus .2ex}{1.6ex plus .2ex}
\titlespacing*{\subsection}{0pt}{2.8ex plus 1ex minus .2ex}{1.3ex plus .2ex}
\titlespacing*{\paragraph}{0pt}{1.8ex plus .5ex minus .2ex}{0.6em}

\newtheoremstyle{slanted}{1.6ex plus .4ex}{1.6ex plus .4ex}{\itshape}{}{\scshape}{.}{0.6em}
  {\thmname{#1}\thmnumber{ #2}\thmnote{ {\normalfont(#3)}}}
\newtheoremstyle{upright}{1.6ex plus .4ex}{1.6ex plus .4ex}{\normalfont}{}{\scshape}{.}{0.6em}
  {\thmname{#1}\thmnumber{ #2}\thmnote{ {\normalfont(#3)}}}
\theoremstyle{slanted}
\newtheorem{theorem}{Theorem}
\newtheorem{proposition}{Proposition}
\newtheorem{lemma}{Lemma}
\newtheorem{corollary}{Corollary}
\theoremstyle{upright}
\newtheorem{definition}{Definition}
\newtheorem{example}{Example}
\newtheorem{assumption}{Assumption}
\newtheorem{remark}{Remark}

\makeatletter
\renewenvironment{proof}[1][\proofname]{\par
  \pushQED{\qed}\normalfont\topsep1.4ex\@plus.4ex\relax
  \trivlist\item[\hskip\labelsep\scshape #1\@addpunct{.}]\ignorespaces}
  {\popQED\endtrivlist\@endpefalse}
\renewenvironment{abstract}{\list{}{\leftmargin0.45in\rightmargin0.45in\topsep0pt}\item\relax}{\endlist}
\makeatother

\newcommand{\E}{\mathbb E}
\newcommand{\R}{\mathbb R}
\newcommand{\cK}{K_{\Pi}}
\newcommand{\calB}{\mathcal B}

\newcommand{\com}{comonotonicity}
\newcommand{\Com}{Comonotonicity}

\title{\large\textbf{Lone Wolves and Lattices in Bayesian Matching with Transfers}}

\author{\normalsize\textsc{Xiyu Li}\thanks{Department of Economics, Georgetown University. Preliminary working paper. This version: November 2026.}\\[0.2em]
\small\texttt{xl674@georgetown.edu}}
\date{}

\begin{document}
\maketitle
\thispagestyle{empty}

\begin{abstract}
I study the \emph{Lone Wolf} and \emph{lattice} properties of stable outcomes in one-to-one matching with transferable utility and one-sided incomplete information. The solution concept is the Bayesian stable market state of \citet{liu2020stability} and \citet{chenhu2024bayesian}. With transfers, the wage in a blocking offer screens the worker's types, changing both whom the firm attracts and what it learns. I show by example that the firm-side and matched-set versions of Lone Wolf, as well as the lattice property, can fail even when outcomes are compared under the same information. I then assume that each worker's matching value is additively separable between his type and the firm's identity. Under this condition, a firm that is unmatched in some outcome of a common-information stratum earns zero in every outcome of that stratum. Workers can retain information rents, so worker-side payoff Lone Wolf can fail. Nevertheless, for fixed worker values, prior, type profile, and information structure, outside a finite union of affine hyperplanes in the space of firm values, all stable outcomes in a stratum leave the same agents unmatched. Whenever this matched-set property holds, the stable payoff vectors at the fixed type profile form a lattice, and worker-optimal and firm-optimal outcomes exist. The main results extend to comonotonicity of relative preferences under posterior independence across workers' types; without this informational condition, the firm-side property can fail.

\medskip
\noindent\textsc{Keywords.} Matching, transfers, incomplete information, Bayesian stability, Lone Wolf, lattice.

\noindent\textsc{JEL classification.} C78, D82, D83.
\end{abstract}

\section{Introduction}
Who remains unmatched in a stable matching? A central result in matching theory,
the \emph{Lone Wolf Theorem}, states that an agent who is unmatched at one stable
matching is unmatched at every stable matching
\citep{mcvitiewilson1970stable,roth1986rural}. Thus, although stability generally
does not determine a unique assignment, it does determine the set of agents who
participate in the market. The result has important practical implications. For
example, a rural hospital that is unable to fill a position under one stable
outcome cannot solve that problem merely by selecting a different stable outcome.

A second central result describes how stable outcomes are ordered. The
\emph{Lattice Theorem} states that if each worker is given the partner he prefers
among his partners at two stable matchings, the result is again a stable
matching, and so is the result of giving each worker the partner he likes less
\citep{knuth1976mariages}. The stable matchings therefore form a lattice. There
are a worker-optimal and a firm-optimal stable matching, and the two sides of
the market have opposed interests over stable matchings.

With transferable utility, stability concerns both a matching and the division of
surplus, and both results become statements about payoffs. The natural Lone Wolf
property is that an agent who is unmatched at some stable outcome earns zero at
every stable outcome. The lattice property is that the stable payoff vectors
form a lattice. Both hold in the assignment game
\citep{shapleyshubik1971assignment,demangegale1985strategy}. This paper asks
whether they survive when workers have private information and firms can use
wages both to divide surplus and to screen workers.

Private information changes the nature of a blocking deviation. When a firm offers
a wage to a worker, acceptance depends on the worker's private type and on his
current wage. Consequently, the set of types who accept an offer is endogenous to
the wage. A cheap offer may attract only those types that the firm least wants to
hire, while attracting every type may require an unprofitably high wage. A wage
is therefore not merely a transfer: it is also a filter. This screening role can
generate the kind of adverse selection familiar from \citep{akerlof1970lemons}.
Acceptance can also be news about the worker whom the firm currently employs,
when the two workers' types are correlated, and the firm re-evaluates him
accordingly.

I study Bayesian stable states in a one-to-one matching market with transfers.
Each worker privately observes his type, while firms may have heterogeneous
information about the type profile. Two stable states can differ simply because
firms know different things. Following \citet{hu2026structure}, I therefore compare
outcomes that share the true type profile and the firms' information partitions.
I call the set of Bayesian stable outcomes with a given type profile and a given
information structure a \emph{stratum}.

The first result is that both properties fail in general. I construct a
one-worker, two-firm, two-type example in which the worker is employed by
different firms in two outcomes of the same stratum. Each firm
earns a strictly positive payoff when it employs the worker and zero when it
does not, and the stable payoffs do not form a lattice. The mechanism is
\emph{bilateral adverse selection}. When the worker is employed by one firm, the
other firm's cheapest offer attracts only a type that this firm does not want,
and attracting both types is too expensive. The same logic operates in both
directions and sustains both states simultaneously.

The main analysis isolates a restriction on values that eliminates this
screening. Worker values are \emph{additively separable} if
\[
a_{ij}(t_i)=g_i(t_i)+h_{ij}.
\]
Under this assumption, private information shifts the level of a worker's utility
but does not change how he compares firms. For an employed worker, the wage at
which he accepts an outside offer is therefore independent of his type. A
deviating firm must attract either all of his types or none; it cannot use the
wage to select a subset.

Two observations drive the analysis under separability. First, a
\emph{rehiring} argument shows that if worker $i$ works for firm $j$ in one stable
state but is unemployed in another, then $j$'s payoff cannot be lower in the
second state: the firm can approximately recreate the old contract. Second, a
\emph{pairwise-opposition} argument shows that if $i$ is employed by another firm
in the second state, then the sum of $i$'s and $j$'s payoffs there cannot fall
below their interim pair surplus. Pairwise opposition need not hold within an
arbitrary state, but it holds between any two outcomes of the same
stratum.

These observations yield the first main theorem: under separability, a firm that
is unmatched in some outcome of a stratum earns zero in every outcome of
that stratum.

Separability cannot simply be dropped. In a one-worker, two-firm, two-type
environment, a failure of separability means that the worker's two types differ
in how strongly they prefer one firm to the other. I show that, whenever this
happens, there is a nonempty open set of firm values for which the
bilateral-adverse-selection construction works. The firm-side failure is
therefore robust, and not a knife-edge phenomenon.

The worker side requires a slightly different answer. A worker who is unemployed in one
stable state can earn a strictly positive payoff in another, even under
separability. Separability eliminates the screening of employed workers but not
the screening of unemployed workers, and a wage that every type of a worker
accepts leaves some types an \emph{information rent}. This failure is
nevertheless non-generic. The second main theorem shows that, once the stratum,
the prior, and separable worker values are fixed, there is a finite union of
affine hyperplanes in the space of firm values outside which all Bayesian stable states
match the same workers and the same firms.

The lattice property follows almost the same pattern. Under separability, any two
outcomes of a stratum that leave the same agents unmatched have a join and a
meet in the stratum: the payoff vector that gives every worker the larger and
every firm the smaller of their two payoffs is attained by a Bayesian stable
outcome, and so is the opposite one. Consequently the stable payoffs form a
lattice, and worker-optimal and firm-optimal outcomes exist, whenever all
outcomes of the stratum leave the same agents unmatched, which is the generic
case. When two outcomes employ different workers, joins and meets can fail to
exist, again because of information rents.

Additive separability restricts workers' values alone. The lesson of
Example~\ref{ex:failure}, however, is that the classical properties fail when
the firms and the different types of a worker have opposed interests: the type
that is most eager to join a firm is the type that the firm values least. The
last part of the paper shows that what ultimately matters is a joint property of workers' and firms'
values. Values exhibit \emph{comonotonicity of relative preferences} if a type
that is relatively more eager to work for a firm, compared with any rival firm,
is at least as valuable to that firm; I say \emph{\com} for short. \Com{} contains additive separability and
private values for firms as special cases, and it also allows workers' rankings
of firms to depend on their types. Under it an offer to an employed worker can
again select among his types, so the firm that makes the offer re-evaluates its
own current worker on the event that the offer is accepted. Under posterior
independence across workers this re-evaluation is vacuous, and the firm-side
theorem, the generic Lone Wolf theorem and the lattice theorem all continue to
hold. Posterior independence cannot be dropped. With correlated types, a worker's willingness
to move can be good news about the worker whom the firm would give up, and I
give an example in which the firm-side theorem fails although values are
comonotone.

The remainder of this section reviews the literature.
Section~\ref{sec:model} introduces the model, Bayesian stability, and the two
structural properties. Section~\ref{sec:failure} presents the general failure
and its adverse-selection mechanism. Section~\ref{sec:separable} contains the
main results under additive separability: the firm-side theorem, generic Lone
Wolf, and the lattice property. Section~\ref{sec:com} extends them to \com{}
under posterior independence across workers. Section~\ref{sec:ntu} compares the model with the one
without transfers. Section~\ref{sec:conclusion} concludes. Proofs are in
Appendix~\ref{app:proofs}.

\paragraph{Related literature.}
This paper connects two strands of literature. The first concerns the structure
of stable outcomes under complete information. \citet{mcvitiewilson1970stable}
prove the Lone Wolf Theorem for one-to-one matching without transfers, and
\citet{roth1986rural} extends it to many-to-one matching as the Rural Hospital
Theorem; \citet{knuth1976mariages} states the Lattice Theorem for such markets. With transferable utility,
\citet{shapleyshubik1971assignment} show that the stable payoffs of the assignment
game form a lattice, and \citet{demangegale1985strategy} establish both properties
in a general model with transfers. \citet{jagadeesan2020lone} extend the Lone Wolf
Theorem to exchange economies.

The second strand studies stability under incomplete information, including
\citep{liu2014stable,liu2020stability,chenhu2024bayesian}. My
solution concept follows the Bayesian stability approach of
\citep{liu2020stability,chenhu2024bayesian}. Two papers study structural
properties in this setting. \citet{hu2026structure} studies the Lone Wolf and
lattice properties in a Bayesian matching model without transfers and introduces
the comparison within a common information structure that I adopt;
Remark~\ref{rem:ntu} discusses how his Lone Wolf result relates to the examples
in this paper. He names the extension to transferable utility as an open
direction. \citet{peralta2025lone} studies the Lone Wolf property with
transferable utility under the belief-free approach of \citep{liu2014stable}. This
paper studies both properties in a Bayesian matching model with transfers.

\section{The matching market}\label{sec:model}

I adopt the setup of \citep{liu2020stability}, in the partitional formulation of
\citep{chenhu2024bayesian}.

Let $I=\{1,\dots,n\}$ be a set of workers and $J=\{n+1,\dots,n+m\}$ be a set of
firms. Let $T_i$ be a finite set of types for worker $i\in I$. Worker $i$'s type
$t_i\in T_i$ is his private information. Denote a type profile by
$t=(t_1,\dots,t_n)\in T:=\times_{i=1}^{n}T_i$. There is a common prior
$\beta\in\Delta(T)$ on type profiles, and $\beta$ has full support. The
characteristics of each firm $j\in J$ are commonly known and are summarized by
the index $j$.

Let $a_{ij}(t_i)\in\R$ and $b_{ij}(t_i)\in\R$ be the \emph{matching values} that
worker $i$ of type $t_i$ and firm $j$ receive, respectively, when they match. For
a profile $t=(t_i,t_{-i})\in T$, I write $a_{ij}(t)\equiv a_{ij}(t_i)$ and
$b_{ij}(t)\equiv b_{ij}(t_i)$. The value of being unmatched is normalized to
zero. Firms have interdependent values: a firm's value from a worker depends on
the worker's private type. Utility is quasi-linear in money. If worker $i$ works
for firm $j$ at the wage $p\in\R$, the worker's payoff is $a_{ij}(t_i)+p$ and the
firm's payoff is $b_{ij}(t_i)-p$. Denote a matching market by
$\Gamma=(I,J,T,\beta,a,b)$.

With continuous wages indifference cannot be excluded, so I do not assume
strict preferences. I impose no assumption on $a$ and $b$.

\subsection{Market states}

A \emph{matching} is a one-to-one function $\mu:I\cup J\to I\cup J$ such that the
following conditions hold for each $i\in I$ and each $j\in J$:
\begin{enumerate}[label=(\roman*)]
\item $\mu(i)\in J\cup\{i\}$,
\item $\mu(j)\in I\cup\{j\}$, and
\item $\mu(i)=j$ if and only if $\mu(j)=i$.
\end{enumerate}
If $\mu(i)=i$ or $\mu(j)=j$, the agent is \emph{unmatched}. A \emph{wage
profile} for $\mu$ is a list $p=(p_i)$ that specifies a wage $p_i\in\R$ for each
matched worker $i$. The pair $(\mu,p)$ is an \emph{outcome}, and it is
publicly observable.

The firms' \emph{information structure} is a partition profile
$\Pi=(\Pi_{n+1},\dots,\Pi_{n+m})$, where for each firm $j\in J$, $\Pi_j$ is a
partition of $T$; $t'\in\Pi_j(t)$ means that if the true type profile is $t$, firm
$j$ cannot distinguish $t$ from $t'$. The partition profile $\Pi$ is publicly observable. Each firm $j$ also observes
which cell $\Pi_j(t)$ contains the realized type profile; the other agents need
not observe that cell. The special case of complete information is denoted by
$\Pi^{CI}$, where $\Pi^{CI}_j(t')=\{t'\}$ for all $t'\in T$ and all $j\in J$. The
\emph{common knowledge partition among firms} $\cK$ is the meet of the firms'
partitions, that is, the finest partition of $T$ that is coarser than every
$\Pi_j$; the cell $\cK(t)$ is the common knowledge event among firms at $t$.

A \textbf{market state}, or simply a \textbf{state}, $(\mu,p,t,\Pi)$ specifies an
outcome $(\mu,p)$, a realized type profile $t$, and an information structure $\Pi$.

For a nonempty event $E\subseteq T$, let $\E[b_{ij}\mid E]$ denote the expectation
of $b_{ij}$ under $\beta(\cdot\mid E)$. Fix a state $(\mu,p,t,\Pi)$. The
\emph{payoffs} at a type profile $t'\in T$ are
\begin{align*}
u_i(t')&:=
\begin{cases}
a_{i\mu(i)}(t')+p_i & \text{if $i$ is matched,}\\
0 & \text{otherwise,}
\end{cases}
\\[4pt]
v_j(t')&:=
\begin{cases}
\E\bigl[b_{\mu(j)j}\mid\Pi_j(t')\bigr]-p_{\mu(j)} & \text{if $j$ is matched,}\\
0 & \text{otherwise.}
\end{cases}
\end{align*}
The \emph{interim pair surplus} of worker $i$ and firm $j$ at $t'$ is
$S_{ij}(t'):=a_{ij}(t')+\E[b_{ij}\mid\Pi_j(t')]$, so that
$u_i(t')+v_j(t')=S_{ij}(t')$ whenever $\mu(i)=j$.

\subsection{Bayesian stability}

Under complete information, stability means individual rationality and the
absence of blocking pairs. Bayesian stability of a market state has three
requirements: (i) individual rationality, (ii) no blocking pair, and (iii)
information stability.

\begin{definition}
A state $(\mu,p,t,\Pi)$ is \textbf{individually rational} (IR) if payoffs are
nonnegative, i.e.,
\[
u_i(t)\ge 0\ \text{ for all } i\in I
\qquad\text{and}\qquad
v_j(t)\ge 0\ \text{ for all } j\in J.
\]
\end{definition}

To define a blocking pair $(i,j)$ for the state $(\mu,p,t,\Pi)$, where
$\mu(i)\neq j$, I first specify firm $j$'s belief when it offers worker $i$ a wage
$q$. Let $D_{ij}(q)$ be the set of type profiles under which worker $i$ benefits
from rematching with firm $j$ at the wage $q$, i.e.,
\begin{equation}\label{eq:willingness}
D_{ij}(q):=\bigl\{t'\in T:\ a_{ij}(t')+q>u_i(t')\bigr\}.
\end{equation}
I call $D_{ij}(q)$ the \emph{willingness set} of the offer. It depends on the
outcome $(\mu,p)$ through $u_i$. Based on its initial knowledge $\Pi_j(t)$ and
the hypothetical knowledge $D_{ij}(q)$, firm $j$'s belief is
$\beta(\cdot\mid\Pi_j(t)\cap D_{ij}(q))$, the off-path belief of
\citep{liu2020stability}. Without transfers, the willingness set is a single event
$D_{\mu,ij}$, as in \citep{hu2026structure}. With transfers it varies with the
wage: a wage is both a price and a filter. The willingness set restricts worker
$i$'s type only: whether $t'$ belongs to it depends on $t'_i$ alone.

Firm $j$ uses this belief to evaluate the worker it approaches and also the
worker it currently employs. For an event $E\subseteq T$ such that
$\Pi_j(t')\cap E\neq\emptyset$, firm $j$'s \emph{re-evaluated payoff} at $t'$ on
$E$ is
\[
v_j(t'\mid E):=
\begin{cases}
\E\bigl[b_{\mu(j)j}\mid\Pi_j(t')\cap E\bigr]-p_{\mu(j)} & \text{if $j$ is matched,}\\
0 & \text{otherwise,}
\end{cases}
\]
so that $v_j(t')=v_j(t'\mid T)$.

\begin{definition}\label{def:blocked}
A state $(\mu,p,t,\Pi)$ is \textbf{blocked} by $(i,j)$, where $\mu(i)\ne j$, if
there is a wage $q\in\R$ such that
\[
a_{ij}(t)+q>u_i(t)
\qquad\text{and}\qquad
\E\bigl[b_{ij}\mid\Pi_j(t)\cap D_{ij}(q)\bigr]-q>v_j\bigl(t\mid D_{ij}(q)\bigr).
\]
\end{definition}

This is the blocking condition of \citet[Definition~5]{chenhu2024bayesian}.%
\footnote{\citet{liu2020stability} writes the right-hand side as
$\max\{0,v_j(t\mid D_{ij}(q))\}$, so that a firm whose re-evaluated payoff is
negative must also expect a positive payoff from the deviation;
\citet{chenhu2024bayesian} drop the maximum. Every result below holds under
either version: in each block constructed in the proofs, the firm's
re-evaluated payoff is either nonnegative or the same as in a state that is
already blocked.}
The first inequality says that $t\in D_{ij}(q)$, so the conditioning events in
the second are nonempty. On the right-hand side of the second inequality, the
firm values its current match on the event that worker $i$ accepts, exactly as
it values the new match on the left-hand side. This is in general not the
payoff $v_j(t)$ that enters individual rationality. Under $\Pi^{CI}$ the two
coincide, and the two inequalities reduce to
$a_{ij}(t)+b_{ij}(t)>u_i(t)+v_j(t)$, the blocking condition of the assignment game
of \citep{shapleyshubik1971assignment}.

The re-evaluated payoff equals the payoff in three cases that recur below.

\begin{lemma}[Re-evaluation]\label{lem:reeval}
Fix a state $(\mu,p,t,\Pi)$, a firm $j$, a type profile $t'\in T$ and an event
$E\subseteq T$ such that $\Pi_j(t')\cap E\neq\emptyset$. Then
$v_j(t'\mid E)=v_j(t')$ in each of the following cases:
\begin{enumerate}[label=(\roman*)]
\item firm $j$ is unmatched;
\item $\Pi_j(t')\subseteq E$;
\item $b_{\mu(j)j}$ is constant on $\Pi_j(t')$.
\end{enumerate}
\end{lemma}

The three cases are immediate. In case (i) both payoffs are zero. In case
(ii), $\Pi_j(t')\cap E=\Pi_j(t')$. In case (iii) both conditional expectations
equal the constant.

Outside these cases the two payoffs can differ. If the type of worker $i$ is
correlated with the type of firm $j$'s current worker, then learning that $i$ is
willingness to move changes what the firm expects from the worker it would give
up. Section~\ref{sec:com} shows that this matters.

Finally, the absence of individual or pairwise deviations must convey no
information to firms beyond what is already contained in $\Pi$. Define
\[
N_{\mu,p,\Pi}:=\bigl\{t'\in T:\ (\mu,p,t',\Pi)\text{ is IR and not blocked}\bigr\}.
\]
Upon observing that nobody deviates, firms learn that the true type profile lies
in $N_{\mu,p,\Pi}$. This is uninformative exactly when the event is already
common knowledge among firms. 

\begin{definition}
A state $(\mu,p,t,\Pi)$ is \textbf{Bayesian stable} if it satisfies three
requirements:
\begin{enumerate}[label=(\roman*)]
\item $(\mu,p,t,\Pi)$ is individually rational.
\item $(\mu,p,t,\Pi)$ is not blocked.
\item information stability holds, that is, $\cK(t)\subseteq N_{\mu,p,\Pi}$.
\end{enumerate}
\end{definition}

Equivalently, a state is Bayesian stable if and only if, for every
$t'\in\cK(t)$, the state $(\mu,p,t',\Pi)$ is IR and not blocked: it is common
knowledge that the state is IR and not blocked. Requirement (iii) is stated in
its common-knowledge form; see Proposition~1 of \citet{hu2026structure} and
\citet{chenhu2024bayesian} for the equivalence with the formulation in terms of
information refinement.

\subsection{Common-information subsets and structural properties}\label{sec:properties}

For each $t\in T$, let
\[
S^{B}(t):=\bigl\{(\mu,p):\ (\mu,p,t,\Pi)\text{ is Bayesian stable for some }\Pi\bigr\}
\]
denote the set of Bayesian stable outcomes at $t$. Under $\Pi^{CI}$, information
stability holds automatically and Bayesian stability reduces to stability in the
assignment game at $t$. Hence $S^{B}(t)\neq\emptyset$ for every $t\in T$ by
\citep{shapleyshubik1971assignment}; see also \citep{liu2020stability,chenhu2024bayesian}.

Two Bayesian stable states may differ because firms know different things, and
then different agents may well be matched. Following \citet{hu2026structure}, I
therefore decompose $S^{B}(t)$ by the information structure that supports each
outcome. For every $t$ and every $\Pi$, define the \emph{common-information
subset}
\begin{equation}\label{eq:stratum}
S^{B}(t,\Pi):=\bigl\{(\mu,p):\ (\mu,p,t,\Pi)\text{ is Bayesian stable}\bigr\}.
\end{equation}
For brevity I call $S^{B}(t,\Pi)$ the \textbf{stratum} at $(t,\Pi)$. Then
$S^{B}(t)$ is the union of the strata $S^{B}(t,\Pi)$ over all information
structures. Existence guarantees $S^{B}(t)\neq\emptyset$, but not
$S^{B}(t,\Pi)\neq\emptyset$ for every $\Pi$. All results below compare outcomes
within one stratum.

I study two properties of a stratum $S^{B}(t,\Pi)$.

\paragraph{Lone Wolf.} With transfers, the natural Lone Wolf property concerns
payoffs. I state it separately for the two sides of the market, because the two
sides will behave differently.

\begin{definition}\label{def:LW}
Fix $t$ and $\Pi$.
\begin{enumerate}[label=(\roman*)]
\item \textbf{Firm-side payoff Lone Wolf} holds on $S^{B}(t,\Pi)$ if every firm
that is unmatched in some outcome in $S^{B}(t,\Pi)$ has payoff $v_j(t)=0$ in every
outcome in $S^{B}(t,\Pi)$.
\item \textbf{Worker-side payoff Lone Wolf} holds on $S^{B}(t,\Pi)$ if every worker
who is unmatched in some outcome in $S^{B}(t,\Pi)$ has payoff $u_i(t)=0$ in every
outcome in $S^{B}(t,\Pi)$.
\item \textbf{Matched-set Lone Wolf} holds on $S^{B}(t,\Pi)$ if the set of
unmatched agents is the same in all outcomes in $S^{B}(t,\Pi)$.
\end{enumerate}
\end{definition}

Matched-set Lone Wolf is the form the property takes in models without
transfers, and it implies both payoff properties. The converse fails already
under complete information, where two stable outcomes may match different
agents as long as those agents earn zero in both.

\paragraph{Lattice.} Consider two outcomes in $S^{B}(t,\Pi)$ with payoffs
$(u,v)$ and $(\hat u,\hat v)$.

\begin{definition}\label{def:lattice}
Fix $t$ and $\Pi$.
\begin{enumerate}[label=(\roman*)]
\item A \textbf{join} of the two outcomes is an outcome in $S^{B}(t,\Pi)$ whose
payoffs at $t$ are $\max\{u_i(t),\hat u_i(t)\}$ for every worker $i$ and
$\min\{v_j(t),\hat v_j(t)\}$ for every firm $j$.
\item A \textbf{meet} of the two outcomes is an outcome in $S^{B}(t,\Pi)$ whose
payoffs at $t$ are $\min\{u_i(t),\hat u_i(t)\}$ for every worker $i$ and
$\max\{v_j(t),\hat v_j(t)\}$ for every firm $j$.
\item The \textbf{lattice property} holds on $S^{B}(t,\Pi)$ if every two outcomes
in $S^{B}(t,\Pi)$ have a join and a meet.
\end{enumerate}
\end{definition}

The join is the best of the two outcomes for every worker and the worst for
every firm, and the meet is the reverse. When the lattice property holds, the
payoff vectors at $t$ of the outcomes in $S^{B}(t,\Pi)$ form a lattice under the
order in which one vector is above another if it gives every worker weakly more
and every firm weakly less. An outcome is \emph{worker-optimal} in $S^{B}(t,\Pi)$
if it gives every worker at least as much, and every firm at most as much, as
any other outcome in $S^{B}(t,\Pi)$; a \emph{firm-optimal} outcome is defined
symmetrically.

\section{Failure of the Lone Wolf and lattice properties}\label{sec:failure}

This section shows that fixing the type profile and the information structure
delivers neither the Lone Wolf property nor the lattice property when wages are
available. I begin with the complete-information benchmark, where both hold.

Under $\Pi^{CI}$ a state $(\mu,p,t,\Pi^{CI})$ is Bayesian stable if and only if
$(\mu,p)$ is a stable outcome of the assignment game with surplus
$\alpha_{ij}:=a_{ij}(t)+b_{ij}(t)$: payoffs are nonnegative, and
$u_i(t)+v_j(t)\ge\alpha_{ij}$ for every worker $i$ and every firm $j$.

\begin{proposition}[Complete information]\label{prop:CI}
For every $t\in T$, firm-side and worker-side payoff Lone Wolf hold on
$S^{B}(t,\Pi^{CI})$.
\end{proposition}

The proof rests on one fact: every stable outcome maximizes total surplus, and
stable payoffs exhaust it. This is the duality of
\citep{shapleyshubik1971assignment}. The same structure delivers the lattice
property.

\begin{proposition}[\citealp{shapleyshubik1971assignment}]\label{prop:CIlattice}
For every $t\in T$, the lattice property holds on $S^{B}(t,\Pi^{CI})$.
\end{proposition}

The example below shows that neither proposition has a counterpart under
incomplete information.

\begin{example}[Bilateral adverse selection]\label{ex:failure}
There is one worker and there are two firms, labelled $A$ and $B$. The worker
has two equally likely types, $t^0$ and $t^1$. Both firms are uninformed:
$\Pi_A=\Pi_B=\{\{t^0,t^1\}\}$. The matching values are
\[
\begin{array}{c|cc|cc}
 & a_{A} & a_{B} & b_{A} & b_{B}\\ \hline
t^0 & 6 & 8 & \phantom{-}8 & -1\\
t^1 & 8 & 1 & -3 & 10
\end{array}
\]
where the worker's index is omitted. Thus $\E[b_A]=\tfrac52$ and
$\E[b_B]=\tfrac92$. Type $t^0$ prefers $B$ to $A$ by $2$ at equal wages, and type
$t^1$ prefers $A$ to $B$ by $7$. Firm $A$ gains from type $t^0$ and loses from
type $t^1$; firm $B$ gains from type $t^1$ and loses from type $t^0$.
\end{example}

Since both firms are uninformed, $\cK(t)=T$ for both type profiles, and a state
is Bayesian stable if and only if it is IR and not blocked at $t^0$ and at $t^1$.
In particular, the stratum is the same at both type profiles. Write
$\Pi^{0}$ for the uninformed structure.

\begin{proposition}\label{prop:failure}
In Example~\ref{ex:failure}, for each $t\in\{t^0,t^1\}$, the stratum
$S^{B}(t,\Pi^{0})$ consists of exactly the following outcomes:
\begin{enumerate}[label=(\alph*)]
\item the worker is matched with $A$ at a wage $p\in[1,\tfrac52]$;
\item the worker is matched with $B$ at a wage $p\in[4,\tfrac92]$.
\end{enumerate}
In the outcomes in (a), firm $A$ earns $\tfrac52-p$ and firm $B$ is unmatched. In
the outcomes in (b), firm $B$ earns $\tfrac92-p$ and firm $A$ is unmatched.
\end{proposition}

\begin{corollary}\label{cor:failure}
In Example~\ref{ex:failure}, the following properties fail on $S^{B}(t,\Pi^{0})$,
for each $t\in\{t^0,t^1\}$.
\begin{enumerate}[label=(\roman*)]
\item Matched-set Lone Wolf and firm-side payoff Lone Wolf: firm $A$ is unmatched
in the outcomes in (b) and earns $\tfrac52-p>0$ in the outcomes in (a) with
$p<\tfrac52$, and symmetrically for firm $B$.
\item The lattice property: the outcome in (a) with wage $1$ and the outcome in
(b) with wage $4$ have neither a join nor a meet.
\end{enumerate}
\end{corollary}

Worker-side payoff Lone Wolf holds in the example, but only because the worker
is matched in every outcome of the stratum.

The two families of outcomes also differ in surplus. At $t^1$, the interim pair
surplus is $S_{A}(t^1)=8+\tfrac52=\tfrac{21}{2}$ with firm $A$ and
$S_{B}(t^1)=1+\tfrac92=\tfrac{11}{2}$ with firm $B$. Bayesian stable outcomes in
one stratum therefore need not maximize the total surplus, which is why the argument behind
Proposition~\ref{prop:CI} is unavailable.

The mechanism behind Example~\ref{ex:failure} is \emph{bilateral adverse
selection}.
Consider a worker $i$ who is matched with firm $k$ at the wage $p_i$, and a firm
$j\ne k$ that offers him the wage $q$. Type $t_i$ accepts if and only if
\[
q>p_i+a_{ik}(t_i)-a_{ij}(t_i).
\]
The types that accept first, as $q$ rises, are those with the largest
\emph{relative} value $a_{ij}(t_i)-a_{ik}(t_i)$ for the new firm. A deviating
firm thus faces a menu of willingness sets, ordered by inclusion, and it can choose
among them by choosing the wage. It has two kinds of offers.

A \emph{cheap offer} attracts only the types that are most eager to move. In
Example~\ref{ex:failure}, when the worker is with $A$, the type most eager to move
to $B$ is $t^0$, the type that $B$ does not want. When the worker is with $B$, the
type most eager to move to $A$ is $t^1$, the type that $A$ does not want. Each type
is most eager to join the firm that values it least. Acceptance of a cheap offer
is bad news for the firm that makes it, as in \citep{akerlof1970lemons}.

A \emph{pooling offer} attracts every type, so acceptance carries no news. But
it must compensate the most reluctant type. In the example, firm $B$ must pay
more than $p+7$ to attract type $t^1$ away from $A$, and firm $A$ must pay more
than $p+2$ to attract type $t^0$ away from $B$. At the wages in
Proposition~\ref{prop:failure}, both exceed the expected value of the worker.

Adverse selection therefore operates in both directions at once. It protects
firm $A$'s match against firm $B$ and firm $B$'s match against firm $A$, and so it
sustains both matchings in the same stratum. Two features of the example are
responsible. First, the worker's ranking of the two firms depends on his type,
so that the willingness set of a cheap offer is a proper subset of the firm's
information cell. Second, firms' values are aligned against that selection.
Section~\ref{sec:separable} removes the first feature: under additive separability,
a worker's ranking of firms does not depend on his type, a deviating firm must
attract all types of an employed worker or none, and job-to-job offers cannot
screen. Section~\ref{sec:com} shows that removing the second feature is enough
under posterior independence across workers.

\begin{remark}[Without transfers]\label{rem:ntu}
Example~\ref{ex:failure} has a counterpart without transfers. Consider the model
of \citet{hu2026structure}: there are no wages, and a worker $i$ and a firm $j$
with $\mu(i)\ne j$ block at $t$ if $a_{ij}(t)>a_{i\mu(i)}(t)$ and
$\E[b_{ij}\mid\Pi_j(t)\cap D_{\mu,ij}]>\E[b_{\mu(j)j}\mid\Pi_j(t)\cap D_{\mu,ij}]$,
where $D_{\mu,ij}=\{t'\in T:a_{ij}(t')>a_{i\mu(i)}(t')\}$. Take the values of
Example~\ref{ex:failure} and the uninformed structure $\Pi^{0}$. If the worker is
matched with $A$, the only type willingness to move to $B$ is $t^0$, and
$b_B(t^0)=-1<0$. If the worker is matched with $B$, the only type willingness to move
to $A$ is $t^1$, and $b_A(t^1)=-3<0$. Both states are individually rational at
both types. Hence both are Bayesian stable under the same $t$ and the same
$\Pi$, with firm $B$ unmatched in the first and firm $A$ unmatched in the second.

Thus common information alone does not ensure that the same agents are
unmatched across Bayesian stable matchings, with or without transfers. The
alternating-chain argument used in \citet{hu2026structure} requires, at each
step, that a worker's willingness to leave his employer for a firm not reveal
payoff-relevant information to that firm. This is so if the worker's ranking of
the two firms is the same at every type profile in $\cK(t)$, as under additive
separability. It is also so if every matched firm observes the type of its own
worker, because along the chain a firm approaches a worker whom it employs in
the other matching and then knows his type. The example has neither feature:
the worker's ranking depends on his type, and the firms are uninformed. The
source of the failure here is the type-dependence of workers' rankings, not
transfers.
Section~\ref{sec:ntu} returns to the model without transfers and gives
conditions on values under which the invariance holds.
\end{remark}

\section{Additive separability}\label{sec:separable}

Adverse selection is what makes both properties fail in
Example~\ref{ex:failure}. A natural response is to restrict preferences so that
it cannot arise. This section adopts a condition that is standard in models of
adverse selection and is used by \citet{liu2020stability} and
\citet{hu2026structure}. Section~\ref{sec:firms} shows that it delivers the
Lone Wolf property for firms, Section~\ref{sec:workers} that it delivers the
property for workers generically, and Section~\ref{sec:lattice} that it
delivers the lattice property whenever all outcomes match the same agents.

\subsection{The Lone Wolf property for firms}\label{sec:firms}

\begin{assumption}[Additive separability]\label{ass:sep}
For every worker $i\in I$ there are a function $g_i:T_i\to\R$ and numbers
$(h_{ij})_{j\in J}$ such that
\[
a_{ij}(t_i)=g_i(t_i)+h_{ij}\qquad\text{for all } j\in J \text{ and all } t_i\in T_i .
\]
\end{assumption}

 Equivalently, the difference
$a_{ij}(t_i)-a_{i\ell}(t_i)=h_{ij}-h_{i\ell}$ does not depend on $t_i$, for all firms
$j$ and $\ell$. A worker's private information shifts the \emph{level} of his
value from working, for example through his reservation value or his cost of
effort, but it does not change how he compares one firm with another. The
assumption places no restriction on firms' values $b$, on the prior, or on the
information structure. Example~\ref{ex:failure} violates it: there
$a_A-a_B$ equals $-2$ at type $t^0$ and $7$ at type $t^1$.

The role of the assumption is to make job-to-job offers uninformative. Consider
a worker $i$ who is employed by firm $\ell$ at the wage $p_i$, and an offer $q$
from another firm $j$. A type profile $s$ belongs to the willingness set
$D_{ij}(q)$ if and only if $a_{ij}(s)+q>a_{i\ell}(s)+p_i$. Under
Assumption~\ref{ass:sep} this inequality reads $h_{ij}+q>h_{i\ell}+p_i$, which
does not involve $s$.

\begin{lemma}[No screening of employed workers]\label{lem:noscreen}
Suppose Assumption~\ref{ass:sep} holds. Fix a state $(\mu,p,t,\Pi)$, a worker $i$
with $\mu(i)=\ell\in J$, a firm $j\ne\ell$, and a wage $q\in\R$. Then
\[
D_{ij}(q)=
\begin{cases}
T & \text{if } q>p_i+h_{i\ell}-h_{ij},\\
\emptyset & \text{otherwise.}
\end{cases}
\]
\end{lemma}

An offer to an employed worker thus attracts all of his types or none. A firm
that poaches him learns nothing from his acceptance and evaluates him at
$\E[b_{ij}\mid\Pi_j(t)]$. It learns nothing about its own current worker either:
by Lemma~\ref{lem:reeval}(ii), it compares the offer with its payoff $v_j(t)$.
The lemma does not extend to unemployed workers. A
worker who is unmatched accepts the offer $q$ from firm $j$ if and only if
$q>-g_i(t_i)-h_{ij}$, and this threshold depends on his type. Offers to unemployed
workers can still screen, which is the source of the worker-side failure in
Section~\ref{sec:workers}.

Two lemmas drive the analysis. Throughout, fix $t$ and $\Pi$, and let $(\mu,p)$ and
$(\hat\mu,\hat p)$ be two outcomes in the stratum $S^{B}(t,\Pi)$. Denote their
payoffs by $(u,v)$ and $(\hat u,\hat v)$, and the willingness sets and the
re-evaluated payoffs in the state $(\hat\mu,\hat p,t,\Pi)$ by $\hat D_{ij}(q)$
and $\hat v_j(\cdot\mid\cdot)$.

The first lemma does not use Assumption~\ref{ass:sep}. If a firm's employee in
one outcome is unmatched in the other, the firm can recreate the old job there.

\begin{lemma}[Rehiring]\label{lem:rehiring}
If $\mu(i)=j$ and $i$ is unmatched under $\hat\mu$, then $\hat v_j(t)\ge v_j(t)$.
\end{lemma}

In the proof, the firm offers the old wage plus a small premium. The offer is
accepted by every type because the old wage was
individually rational at every type profile that firms could not rule out.
Screening an unemployed worker is possible in general, but the firm does not
need to screen in order to get its old payoff back.

The second lemma is where Assumption~\ref{ass:sep} is used. Under complete
information, stability requires $u_i+v_j\ge\alpha_{ij}$ for every worker and every
firm. Under incomplete information this inequality can fail within a state, as
Example~\ref{ex:failure} shows. The lemma recovers it \emph{across} two outcomes
of a stratum, for a pair that is matched in one of them.

\begin{lemma}[Pairwise opposition]\label{lem:opposition}
Suppose Assumption~\ref{ass:sep} holds. If $\mu(i)=j$ and $\hat\mu(i)=\ell\in J$
with $\ell\neq j$, then
\[
\hat u_i(t)+\hat v_j(t)\ \ge\ S_{ij}(t)\ =\ u_i(t)+v_j(t).
\]
\end{lemma}

In words, a worker and a firm who are matched in one outcome cannot both be
worse off in another outcome in which the worker is employed elsewhere. If they
were, the firm could make every type of the worker indifferent between his new
employer and itself, add a small premium, and hire him at its original belief.

The two lemmas deliver the first main result.

\begin{theorem}\label{thm:firm}
Suppose Assumption~\ref{ass:sep} holds. For every $t$ and every $\Pi$, firm-side
payoff Lone Wolf holds on $S^{B}(t,\Pi)$: a firm that is unmatched in some
outcome in $S^{B}(t,\Pi)$ earns zero in every outcome in $S^{B}(t,\Pi)$.
\end{theorem}

\begin{figure}[!htb]
\centering
\begin{tikzpicture}[x=1.75cm,y=1.25cm,font=\small]
\node at (-0.9,1) {$J$:};
\node at (-0.9,0) {$I$:};
\node (j0) at (0,1) {$j_0$};
\node (i1) at (1,0) {$i_1$};
\node (j1) at (2,1) {$j_1$};
\node (i2) at (3,0) {$i_2$};
\node (j2) at (4,1) {$j_2$};
\node (dd) at (5,0) {$\cdots$};
\draw (j0) -- node[above right=-2pt]{$\mu$} (i1);
\draw[dashed] (i1) -- node[above left=-2pt]{$\hat\mu$} (j1);
\draw (j1) -- node[above right=-2pt]{$\mu$} (i2);
\draw[dashed] (i2) -- node[above left=-2pt]{$\hat\mu$} (j2);
\draw (j2) -- node[above right=-2pt]{$\mu$} (dd);
\node[above=1pt] at (j0.north) {$\Delta v_{j_0}=\gamma$};
\node[below=1pt] at (i1.south) {$\Delta u_{i_1}\ge\gamma$};
\node[above=1pt] at (j1.north) {$\Delta v_{j_1}\ge\gamma$};
\node[below=1pt] at (i2.south) {$\Delta u_{i_2}\ge\gamma$};
\node[above=1pt] at (j2.north) {$\Delta v_{j_2}\ge\gamma$};
\end{tikzpicture}
\caption{The alternating path in the proof of Theorem~\ref{thm:firm}. Firm $j_0$
is unmatched under $\hat\mu$ and earns $\gamma>0$ under $\mu$. Solid edges are
matched under $\mu$ and dashed edges under $\hat\mu$. Moving from $(\mu,p)$ to
$(\hat\mu,\hat p)$, every firm on the path loses at least $\gamma$ and every
worker gains at least $\gamma$.}\label{fig:chain}
\end{figure}
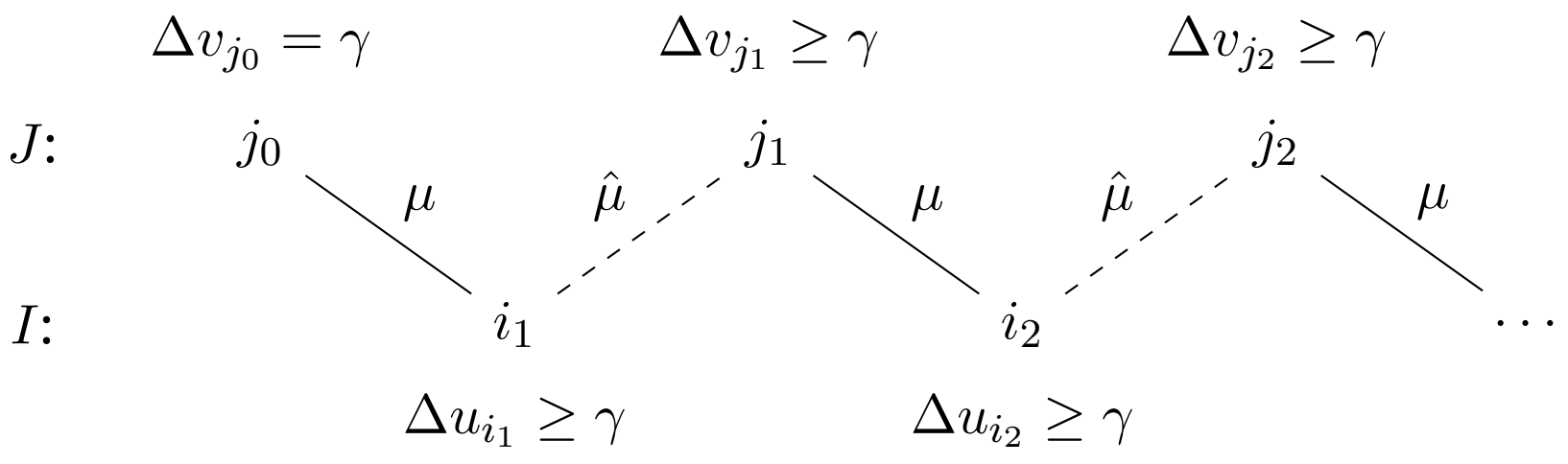

The proof follows the alternating-chain method of the classical Lone Wolf
arguments. Here
the chain carries a payoff gap: each worker on the path gains
at least $\gamma$ in the second outcome and each firm loses at least $\gamma$. The
path cannot end at a worker, by rehiring, and it cannot end at a firm, because
a firm that loses $\gamma$ had a positive payoff to lose.
Figure~\ref{fig:chain} illustrates.

Three remarks delimit the result.

First, Theorem~\ref{thm:firm} concerns firms' payoffs. It does not say that the
same firms are matched in all outcomes of a stratum: a firm may be matched in
one outcome and unmatched in another while earning zero in both, as under
complete information.

Second, the theorem is silent about workers. Lemma~\ref{lem:rehiring} protects
a firm whose employee becomes unemployed, and it has no counterpart for a
worker whose employer becomes idle. Section~\ref{sec:workers} shows that worker-side payoff
Lone Wolf can fail under Assumption~\ref{ass:sep}.

Third, Assumption~\ref{ass:sep} is used only in Lemma~\ref{lem:opposition}, and
only to ensure that a worker's acceptance of a job-to-job offer carries no
news for the firm that makes it: it is not bad news about him, and it is no
news about the firm's current worker. Section~\ref{sec:com} separates the two.

Theorem~\ref{thm:firm} shows that additive separability is sufficient for
firm-side payoff Lone Wolf. It cannot be dropped as long as only workers'
values are restricted: in the smallest environment in which separability can
fail, its failure allows a robust violation. The same conclusion can hold
without it, however, when firms' values are suitably related to workers' values
and posterior beliefs are independent across workers, as Section~\ref{sec:com} shows.

Consider the environment of Example~\ref{ex:failure} with arbitrary values.
There is one worker with two types, $L$ and $H$, and a full-support prior; there
are two firms, $A$ and $B$, both uninformed, so the information structure is
$\Pi^{0}$. Omit the worker's index and write $\delta_s:=a_A(s)-a_B(s)$ for
$s\in\{L,H\}$. Assumption~\ref{ass:sep} holds if and only if $\delta_L=\delta_H$. If
it fails, label the types so that $\delta_L<\delta_H$.

\begin{proposition}[Failure without separability]\label{prop:necessity}
Suppose that $\delta_L<\delta_H$. There is a nonempty open set of firm values
$b\in\R^4$ for which, at each $t\in\{L,H\}$, the stratum $S^{B}(t,\Pi^{0})$ contains
an outcome in which the worker is matched with $A$ and firm $A$ earns a positive
payoff, and an outcome in which the worker is matched with $B$. For these firm
values, firm-side payoff Lone Wolf and matched-set Lone Wolf fail on
$S^{B}(t,\Pi^{0})$.
\end{proposition}

The proof fixes wages $p_A$ and $p_B$ with $\delta_L<p_B-p_A<\delta_H$ and
chooses firm values such that, in each of the two outcomes, the cheap offer of
the idle firm attracts only a type that it does not want, and its pooling offer
is too expensive.
Example~\ref{ex:failure} is an instance of this construction, with $t^0$ in the
role of $L$ and $t^1$ in the role of $H$, and for instance $p_A=2$ and $p_B=\tfrac{17}{4}$.
The construction does not require the two types to rank the firms differently:
$\delta_L$ and $\delta_H$ may have the same sign. It only requires that they differ
in how strongly they prefer one firm to the other. It works by making each
firm's more eager type its less valuable one, since $b_A(H)<\E[b_A]$ and
$b_B(L)<\E[b_B]$. Section~\ref{sec:com} takes this as its starting point.

\subsection{Workers and generic Lone Wolf}\label{sec:workers}

Section~\ref{sec:firms} is about firms. This subsection turns to workers and to
the set of agents who are matched. A worker who is unemployed in one outcome of
a stratum can earn a positive payoff in another, even under
Assumption~\ref{ass:sep}, because he can keep an information rent. However, I show that
such a failure is a knife-edge event: for all firm values outside a finite
union of affine hyperplanes, every outcome of a stratum matches the same agents. On the
exceptional hyperplanes, a worker who is unemployed in one outcome earns exactly
his information rent in any outcome in which he is employed.

The following example shows that the analogue of Theorem~\ref{thm:firm} for
workers is false, even under Assumption~\ref{ass:sep}.

\begin{example}[An information rent]\label{ex:worker}
There are two workers, $1$ and $2$, and two firms, $A$ and $B$. Worker $1$ has
two equally likely types, $t^0$ and $t^1$; worker $2$ has a single type. Both
firms are uninformed. The matching values are
\[
\begin{array}{l|cc|cc}
 & a_{iA} & a_{iB} & b_{iA} & b_{iB} \\ \hline
i=1,\ t^0 & -1 & -1 & 3  & -1 \\
i=1,\ t^1 & -3 & -3 & 0  & \phantom{-}7  \\
i=2       & 15 & \phantom{-}7  & 5  & \phantom{-}1
\end{array}
\]
Worker values are additively separable. The true type of worker $1$ is $t^0$.
Note that $\E[b_{1A}]=\tfrac32$ and $\E[b_{1B}]=3$.
\end{example}

\begin{proposition}\label{prop:worker}
In Example~\ref{ex:worker}, for every $p_2\in[-7,3]$ the stratum contains the
following two outcomes:
\begin{enumerate}[label=(\alph*)]
\item worker $1$ is matched with $B$ at the wage $3$, and worker $2$ is matched
with $A$ at the wage $p_2$;
\item worker $1$ is unmatched, and worker $2$ is matched with $A$ at the wage
$p_2$.
\end{enumerate}
Worker $1$ earns $2$ in (a) and $0$ in (b). Firm $B$ earns $0$ in both. Hence
worker-side payoff Lone Wolf and matched-set Lone Wolf fail.
\end{proposition}

In outcome (a) the wage of worker $1$ is pinned from both sides. It must be at
least $3$ for type $t^1$ to accept the job, and at most $\E[b_{1B}]=3$ for firm
$B$ to offer it. Type $t^0$, who would work for less, keeps an
\emph{information rent} of $2$. Competition does not remove the rent: in outcome
(b), a wage below $3$ is accepted only by type $t^0$, whom firm $B$ values at
$-1$. This is adverse selection again, but among the types of an
\emph{unemployed} worker. Assumption~\ref{ass:sep} does not exclude it:
Lemma~\ref{lem:noscreen} applies to employed workers only.

The example also shows why the complete-information argument breaks down. Under
complete information, a firm could hire an unemployed worker for almost
nothing, so a worker who is unemployed in one stable outcome cannot earn a
positive payoff in another. Here the firm is indifferent between hiring worker
$1$ and not hiring him, and its indifference does not make him indifferent.

The rent in the example can be read off workers' values. For a worker $i$ and
a firm $j$, define
\[
\rho_{ij}(t):=a_{ij}(t)-\min_{s\in\cK(t)}a_{ij}(s)\ \ge\ 0 .
\]
This is the payoff of worker $i$ at $t$ when he works for $j$ at the lowest wage
that is individually rational at every type profile in $\cK(t)$. It depends on
worker values and on the information structure, and not on firm values. It is
zero if firms commonly know worker $i$'s type, or if his true type is the one
that values the job least among those in $\cK(t)$. The next lemma shows that
$\rho_{ij}(t)$ is exactly what a worker earns in the situation of
Lemma~\ref{lem:rehiring} when the firm does not strictly gain from his
unemployment. It needs no assumption.

\begin{lemma}[Pinned wage]\label{lem:pinned}
Let $(\mu,p)$ and $(\hat\mu,\hat p)$ be outcomes in $S^{B}(t,\Pi)$. Suppose that
$\mu(i)=j$, that $i$ is unmatched under $\hat\mu$, and that
$\hat v_j(t)=v_j(t)$. Then $u_i(t)=\rho_{ij}(t)$.
\end{lemma}

I now show that two outcomes of a stratum can match different agents only if
firm values satisfy one of finitely many linear equations. The equations are
indexed by alternating paths.

A \emph{path} is a sequence $c=(k_0,k_1,\dots,k_L)$ of distinct agents, with
$L\ge1$, in which workers and firms alternate. Its \emph{edges} are the pairs
$e_r=\{k_{r-1},k_r\}$, $r=1,\dots,L$; each consists of a worker and a firm, and I
write $S_{e}(t)$ and $\rho_{e}(t)$ for $S_{ij}(t)$ and $\rho_{ij}(t)$ when
$e=\{i,j\}$. Let $P(c)$ be the set of edges with odd index and $\hat P(c)$ the
set of edges with even index. An endpoint of the path that is a worker is a
\emph{worker endpoint}; it belongs to exactly one edge. Define
\[
R_c:=\sum_{\substack{\text{worker endpoints}\\ \text{with edge } e\in P(c)}}\rho_e(t)
\ -\sum_{\substack{\text{worker endpoints}\\ \text{with edge } e\in\hat P(c)}}\rho_e(t),
\]
and
\begin{equation}\label{eq:Fc}
F_c(b):=\sum_{e\in P(c)}S_e(t)-\sum_{e\in\hat P(c)}S_e(t)-R_c .
\end{equation}
Since $S_{ij}(t)=a_{ij}(t)+\E[b_{ij}\mid\Pi_j(t)]$, the function $F_c$ is affine in
$b$. Let
\[
H_c:=\{b:\ F_c(b)=0\}
\qquad\text{and}\qquad
\mathcal H:=\bigcup_{c}H_c ,
\]
where the union is over all paths. There are finitely many paths, and
$\mathcal H$ depends on $t$, $\Pi$, $\beta$ and $a$ only. Each $H_c$ is a
genuine affine hyperplane, not a lower-dimensional affine subset: a path has at
least one edge, and along every edge the interim surplus contains the expected
firm value $\E[b_{ij}\mid\Pi_j(t)]$ with coefficient $1$ or $-1$. These are
nonconstant affine functionals of $b$, so $F_c$ has a nonzero linear part.

For example, on a one-edge path consisting of worker $i$ and firm $j$, oriented
so that its edge belongs to $P(c)$, the equation $F_c(b)=0$ is simply
$S_{ij}(t)=\rho_{ij}(t)$: the pair's interim surplus equals the endpoint
worker's information rent.

When the path has two firm endpoints, $R_c=0$ and $H_c$ is a \emph{surplus tie}:
the pairs in $P(c)$ and the pairs in $\hat P(c)$ generate the same total interim
surplus. This is the counterpart of the ties that make the optimal assignment
non-unique under complete information. When the path has a worker endpoint, the
tie is corrected by that worker's information rent.

For two outcomes $(\mu,p)$ and $(\hat\mu,\hat p)$, consider the graph on
$I\cup J$ whose edges are the pairs that are matched under exactly one of
$\mu$ and $\hat\mu$. Every agent belongs to at most two edges, so each connected
component is a path or a cycle. An agent who is matched under exactly one of the
two matchings belongs to exactly one edge, and is therefore an endpoint of a
component that is a path.

\begin{lemma}[Binding paths]\label{lem:binding}
Suppose Assumption~\ref{ass:sep} holds. Let $(\mu,p)$ and $(\hat\mu,\hat p)$ be
outcomes in $S^{B}(t,\Pi)$, and let $c$ be a path that is a connected component
of the graph defined above. Then $F_c(b)=0$. Moreover, every firm on $c$ earns the
same payoff at $t$ in both outcomes, every worker on $c$ who is matched under
both matchings earns the same payoff at $t$ in both, and a worker on $c$ who is
matched with firm $j$ under exactly one of the matchings earns $\rho_{ij}(t)$ in
that outcome.
\end{lemma}

\begin{theorem}[Generic Lone Wolf]\label{thm:generic}
Suppose Assumption~\ref{ass:sep} holds. Fix $t$, $\Pi$ and $\beta$.
\begin{enumerate}[label=(\roman*)]
\item If $b\notin\mathcal H$, then matched-set Lone Wolf holds on
$S^{B}(t,\Pi)$: all outcomes in $S^{B}(t,\Pi)$ have the same set of unmatched
agents. Consequently, firm-side and worker-side payoff Lone Wolf hold as well.
\item For every path $c$, the set $H_c$ is an affine hyperplane in the space of
firm values. Consequently, $\mathcal H$ is closed, has empty interior, and has
Lebesgue measure zero.
\end{enumerate}
\end{theorem}

Example~\ref{ex:worker} lies on one of the hyperplanes. The path is $(B,1)$: its
single edge is matched in outcome (a) only, and worker $1$ is a worker endpoint.
The equation $F_c(b)=0$ reads $S_{1B}(t^0)=\rho_{1B}(t^0)$, that is,
$-1+\E[b_{1B}]=-1-(-3)$, or $\E[b_{1B}]=3$. Any other value of $\E[b_{1B}]$ removes
one of the two outcomes.

Theorem~\ref{thm:generic} can be read as a statement about perturbations.
Failures of worker-side Lone Wolf under additive separability are fragile:
suitable arbitrarily small changes in firms' values restore matched-set Lone
Wolf. The statement is existential; it does not assert that every small
perturbation restores the property.

\begin{corollary}[Perturbation]\label{cor:perturb}
Suppose Assumption~\ref{ass:sep} holds. Fix $t$, $\Pi$ and $\beta$, and let $b$ be
firm values at which worker-side payoff Lone Wolf fails on $S^{B}(t,\Pi)$. Then
for every $\varepsilon>0$ there are firm values $b'$ with
$\lVert b'-b\rVert<\varepsilon$ at which matched-set Lone Wolf, and hence
worker-side and firm-side payoff Lone Wolf, hold on $S^{B}(t,\Pi)$. This is the
case for every $b'\notin\mathcal H$, and $\mathcal H$ is a closed set of measure
zero that does not depend on $b$.
\end{corollary}

The perturbation changes firms' values only. Workers' values $a$, and with them
Assumption~\ref{ass:sep}, are untouched. The corollary does not say that both
outcomes survive the perturbation with their payoffs aligned. In
Example~\ref{ex:worker}, raising or lowering $\E[b_{1B}]$ slightly removes one of
the two outcomes from the common-information subset. Lone Wolf is restored
because the set of stable outcomes shrinks.

Lemma~\ref{lem:binding} also says what replaces worker-side payoff Lone Wolf on
the exceptional hyperplanes.

\begin{corollary}[A lone wolf earns his information rent]\label{cor:rent}
Suppose Assumption~\ref{ass:sep} holds. If a worker $i$ is unmatched in some
outcome in $S^{B}(t,\Pi)$ and is matched with firm $j$ in another, then in the
latter he earns exactly $\rho_{ij}(t)$, and firm $j$ earns the same payoff at $t$
in both outcomes.
\end{corollary}

Under complete information a lone wolf earns zero. Under incomplete information
with additively separable values, a firm that is a lone wolf still earns zero,
by Theorem~\ref{thm:firm}, while a worker who is a lone wolf earns his
information rent. Worker-side payoff Lone Wolf in the usual sense therefore
holds on a stratum whenever the relevant rents are zero, in particular when
firms commonly know the workers' types, and otherwise it holds for all firm
values outside $\mathcal H$.

Theorem~\ref{thm:generic} also describes the exceptions. Two outcomes of a
stratum can match different agents only when, along the path that connects
them, every firm earns the same payoff in both outcomes, every worker who
changes employer earns the same payoff in both, and every worker who loses his
job was earning exactly his information rent.

\subsection{The lattice property}\label{sec:lattice}

Under complete information the stable payoffs form a lattice
(Proposition~\ref{prop:CIlattice}), and Corollary~\ref{cor:failure} shows that
this fails in Example~\ref{ex:failure}. This subsection asks what additive
separability restores. The answer parallels Section~\ref{sec:workers}. Two
outcomes of a stratum have a join and a meet whenever they leave the same
agents unmatched, which is the generic case by Theorem~\ref{thm:generic}. When
they employ different workers, joins and meets can fail to exist, and the
reason is again an information rent.

\begin{example}[No join and no meet]\label{ex:lattice}
There is one firm and there are two workers. Each worker $i\in\{1,2\}$ has two
equally likely types, $t^0$ and $t^1$, drawn independently, with
\[
a_i(t^0)=-1,\qquad a_i(t^1)=-3,\qquad b_i(t^0)=0,\qquad b_i(t^1)=8 .
\]
The firm is uninformed, and both workers have type $t^0$. With a single firm,
Assumption~\ref{ass:sep} holds trivially.
\end{example}

\begin{proposition}\label{prop:latticefail}
In Example~\ref{ex:lattice}, the stratum $S^{B}(t,\Pi^{0})$ contains the outcome
in which worker $1$ is employed at the wage $3$ and the outcome in which worker
$2$ is employed at the wage $3$. These two outcomes have neither a join nor a
meet.
\end{proposition}

The two outcomes are an instance of Corollary~\ref{cor:rent}. The firm earns the
same payoff in both, and the employed worker earns his information rent
$\rho_{i}(t)=-1-(-3)=2$. The rent goes to whichever worker has the job, and it
cannot be given to both workers or to neither while the firm keeps its payoff.
This is why the join and the meet do not exist. The failure requires two
outcomes with different sets of unmatched agents, and so, by
Theorem~\ref{thm:generic}, firm values on the exceptional set $\mathcal H$. Here
the relevant hyperplane is $\E[b_1]=\E[b_2]$, the one associated with the path
that leads from worker $1$ through the firm to worker $2$: if the two
expectations differ, the firm strictly prefers one of the two workers at the
wage $3$, and the outcome that employs the other worker is blocked.

Now consider outcomes that leave the same agents unmatched.
Fix $t$ and $\Pi$, and let $(\mu,p)$ and $(\hat\mu,\hat p)$ be outcomes in
$S^{B}(t,\Pi)$ with the same set of unmatched agents, with payoffs $(u,v)$ and
$(\hat u,\hat v)$. Consider the graph whose edges are the pairs matched under
exactly one of $\mu$ and $\hat\mu$. No agent is matched under exactly one of the
two matchings, so every agent belongs to zero or two edges, and every connected
component with an edge is a cycle. Call a \emph{component} either the set of
agents on such a cycle or a pair that is matched under both matchings. The
components partition the set of matched agents, and both $\mu$ and $\hat\mu$
match the agents of each component among themselves.

The next lemma is the main step. Within a component, every worker gains and
every firm loses the same amount when the market moves from one outcome to the
other, and this amount is the same at every type profile that firms consider
possible.

\begin{lemma}[Common differences]\label{lem:common}
Suppose Assumption~\ref{ass:sep} holds, and let $(\mu,p)$ and $(\hat\mu,\hat p)$
be outcomes in $S^{B}(t,\Pi)$ with the same set of unmatched agents. For every
component $C$ there is a number $d_C$ such that, for every $t'\in\cK(t)$,
\begin{align*}
\hat u_i(t')-u_i(t')&=d_C\quad\text{for every worker } i\in C,\\
v_j(t')-\hat v_j(t')&=d_C\quad\text{for every firm } j\in C .
\end{align*}
\end{lemma}

Under complete information the corresponding statement is that the workers and
firms of a component have opposite interests in the comparison of two stable
outcomes. Lemma~\ref{lem:common} adds that the size of the common gain is the
same at every type profile in $\cK(t)$. This is what allows the two outcomes to
be recombined component by component without changing what any firm can infer
from a worker's willingness to move.

\begin{proposition}[Joins and meets]\label{prop:lattice}
Suppose Assumption~\ref{ass:sep} holds. If two outcomes in $S^{B}(t,\Pi)$ have
the same set of unmatched agents, then they have a join and a meet.
\end{proposition}

The join is obtained by recombining the two outcomes: on each component it
follows the outcome that the workers of the component prefer. The meet follows
the outcome that the firms of the component prefer. By
Lemma~\ref{lem:common} the two sides of a component disagree at every type
profile in $\cK(t)$, so both recombinations are well defined, and by
Lemma~\ref{lem:noscreen} a block of a recombined outcome would also block one
of the two original outcomes.

\begin{theorem}[Lattice]\label{thm:lattice}
Suppose Assumption~\ref{ass:sep} holds, and suppose that matched-set Lone Wolf
holds on $S^{B}(t,\Pi)$. By Theorem~\ref{thm:generic}, this is the case for all
$b\notin\mathcal H$. Then the lattice property holds on $S^{B}(t,\Pi)$. If
$S^{B}(t,\Pi)$ is nonempty, it contains a worker-optimal outcome and a
firm-optimal outcome.
\end{theorem}

Theorem~\ref{thm:lattice} and Proposition~\ref{prop:latticefail} together describe
the lattice property under additive separability. Generically all outcomes of
a stratum match the same agents and the stable payoffs form a lattice, as under
complete information. On the exceptional set $\mathcal H$ a stratum may contain
outcomes that employ different workers, and such outcomes need not have a join
or a meet.

\section{Comonotonicity of relative preferences}\label{sec:com}

Additive separability restricts workers' values alone. Example~\ref{ex:failure}
suggests that this is more than is needed: its two stable matchings coexist
because the worker's comparison of the two firms depends on his type \emph{and}
because each firm dislikes the type that is most eager to join it. Section~\ref{sec:separable}
removed the first feature. This section shows that it is enough to remove the
second under posterior independence across workers. I introduce a joint
condition on workers' and firms' values, show that it contains additive
separability, and show that the firm-side theorem, the generic Lone Wolf
theorem, and the lattice theorem of Section~\ref{sec:separable} continue to
hold under it whenever the information structure satisfies posterior
independence. This condition is needed because an offer to an employed worker
can now select among his types, and the firm that makes the offer re-evaluates
its current worker on the selected event. An example shows that the firm-side
theorem fails without it.

\subsection{Comonotonicity and posterior independence}

\begin{assumption}[Comonotonicity of relative preferences]\label{ass:com}
For every worker $i\in I$, all firms $j,\ell\in J$ with $j\ne\ell$, and all
types $s,s'\in T_i$,
\[
a_{ij}(s)-a_{i\ell}(s)>a_{ij}(s')-a_{i\ell}(s')
\quad\text{implies}\quad
b_{ij}(s)\ge b_{ij}(s').
\]
\end{assumption}

I refer to the assumption as \emph{\com} for short. The difference
$a_{ij}(s)-a_{i\ell}(s)$ measures how much type $s$ prefers firm
$j$ to firm $\ell$ at equal wages. Assumption~\ref{ass:com} says that a worker's
relative eagerness to work for a firm and his value to that firm move together
across his types: a type that is keener on $j$, relative to any rival, is at
least as valuable to $j$. The condition restricts neither the level of workers'
values, nor the size or the sign of firms' values, nor the prior, nor the
information structure. It compares firms with each other and is silent about
how a worker's types compare a firm with being unmatched. Three classes of
markets satisfy it.

\emph{Additive separability.} Under Assumption~\ref{ass:sep} the difference
$a_{ij}(s)-a_{i\ell}(s)$ is the same for every type. The hypothesis of
Assumption~\ref{ass:com} is never satisfied, so the assumption holds whatever the
firms' values are.

\emph{Private values.} If $b_{ij}$ does not depend on the worker's type, the
conclusion of Assumption~\ref{ass:com} always holds, whatever the workers' values
are.

\emph{Specialists.} Suppose that each type of worker $i$ is a specialist for one
firm: $T_i=J$, and $a_{ij}(s)=h_{ij}+\gamma_i$ if $s=j$ and $a_{ij}(s)=h_{ij}$
otherwise, where $\gamma_i>0$. Assumption~\ref{ass:com} holds if and only if each
firm values its own specialist at least as much as any other type, and values
all other types equally. Here a worker's comparison of firms depends on his
type, so Assumption~\ref{ass:sep} fails.

Example~\ref{ex:failure} violates the assumption in both directions. Type $t^1$
is relatively more eager to work for $A$, and $b_A(t^1)<b_A(t^0)$; type $t^0$ is
relatively more eager to work for $B$, and $b_B(t^0)<b_B(t^1)$.

With two firms, Assumption~\ref{ass:com} orders a worker's types once, by how
much they prefer the first firm to the second, and asks that the first firm's
values be weakly increasing and the second firm's values weakly decreasing
along this order. With three or more firms
it is more demanding: firm $j$'s values must respect the ordering induced by
each rival $\ell$, and if two types are ordered one way against one rival and
the other way against another, firm $j$ must value them equally.

Under Assumption~\ref{ass:com} an offer to an employed worker may attract some
of his types and not others, so Lemma~\ref{lem:noscreen} no longer holds. If
the firm that makes the offer employs a worker of its own, the blocking
condition compares the offer with the firm's re-evaluated payoff, and none of
the cases of Lemma~\ref{lem:reeval} need apply. The re-evaluation is vacuous
under posterior independence across workers.

\begin{assumption}[Posterior independence across workers]\label{ass:ind}
For every firm $j\in J$ and every cell $C\in\Pi_j$, the posterior belief
$\beta(\cdot\mid C)$ is a product distribution across workers:
\[
\beta(\cdot\mid C)=\prod_{i\in I}\beta_i^{C},
\qquad\text{where }\beta_i^{C}\in\Delta(T_i)\text{ for every }i\in I .
\]
\end{assumption}

This is a joint restriction on the prior and the information structure, rather
than a product-prior assumption alone. A sufficient special case is a product
prior together with product information cells: for every firm $j$ and every
cell $C\in\Pi_j$, $C=\prod_{i\in I}C_i$ for some $C_i\subseteq T_i$. The
assumption is automatic when at most one worker has private information. When
all firms are uninformed, it is equivalent to independence under the prior. A
product prior by itself is not sufficient, because conditioning on a
non-product cell can induce correlation between workers' types.

\begin{lemma}[No news about the incumbent]\label{lem:noincumbent}
Suppose Assumption~\ref{ass:ind} holds. Let $j\in J$ and $t'\in T$, let $i$ and
$i'$ be distinct workers, and let $E\subseteq T$ be an event that restricts
only worker $i$'s type, that is, $E=E_i\times T_{-i}$ for some
$E_i\subseteq T_i$. If $\Pi_j(t')\cap E\neq\emptyset$, then
\[
\E\bigl[b_{i'j}\mid\Pi_j(t')\cap E\bigr]=\E\bigl[b_{i'j}\mid\Pi_j(t')\bigr].
\]
Consequently, in every state $(\mu,p,t,\Pi)$, for every worker $i$, every firm
$j\neq\mu(i)$ and every wage $q$ such that
$\Pi_j(t')\cap D_{ij}(q)\neq\emptyset$,
\[
v_j\bigl(t'\mid D_{ij}(q)\bigr)=v_j(t').
\]
\end{lemma}

Under Assumption~\ref{ass:ind}, therefore, a state is blocked if and only if it
is blocked when the firm's re-evaluated payoff $v_j(t\mid D_{ij}(q))$ is
replaced by its payoff $v_j(t)$. Assumption~\ref{ass:ind} plays no role in
Section~\ref{sec:separable}: under additive separability an
offer to an employed worker has the willingness set $T$, and the offers to
unemployed workers that the proofs construct are accepted at every type profile in
$\cK(t)$. Nor is it needed when all firms have private values, by
Lemma~\ref{lem:reeval}(iii).

\Com{} enters the analysis through one property: a worker's willingness
to leave his employer for firm $j$ is never bad news about him for $j$.

\begin{lemma}[Selection]\label{lem:selection}
Suppose Assumption~\ref{ass:com} holds. Let $i\in I$, let $j,\ell\in J$ with
$j\neq\ell$, let $c\in\R$, and let
$U:=\{s\in T:\ a_{ij}(s)-a_{i\ell}(s)\ge c\}$. For every event $E\subseteq T$ such
that $E\cap U\ne\emptyset$,
\[
\E\bigl[b_{ij}\mid E\cap U\bigr]\ \ge\ \E\bigl[b_{ij}\mid E\bigr].
\]
\end{lemma}

Lemma~\ref{lem:rehiring} does not use Assumption~\ref{ass:sep} and is unaffected.
Lemma~\ref{lem:opposition} survives under Assumptions~\ref{ass:com}
and~\ref{ass:ind}, with a different offer in its proof. As in
Section~\ref{sec:separable}, fix $t$ and $\Pi$ and two outcomes $(\mu,p)$ and
$(\hat\mu,\hat p)$ in $S^{B}(t,\Pi)$.

\begin{lemma}[Pairwise opposition under \com]\label{lem:oppcom}
Suppose Assumptions~\ref{ass:com} and~\ref{ass:ind} hold. If $\mu(i)=j$ and $\hat\mu(i)=\ell\in J$
with $\ell\neq j$, then
\[
\hat u_i(t)+\hat v_j(t)\ \ge\ S_{ij}(t)\ =\ u_i(t)+v_j(t).
\]
\end{lemma}

The proof follows the proof of Lemma~\ref{lem:opposition} with a different
offer. The offer is the cheapest one that the worker's true type accepts.
It attracts the types that are at least as eager as the true type to move from
$\ell$ to $j$. Under additive separability every type is equally eager, the
offer attracts all of them, and acceptance is uninformative. Under
Assumption~\ref{ass:com} acceptance may be informative, but it is good news
about the worker who accepts, and under Assumption~\ref{ass:ind} it is no news
about the worker whom the firm would give up.

\subsection{Results}

\begin{theorem}[Firm-side payoff Lone Wolf]\label{thm:firmcom}
Suppose Assumption~\ref{ass:com} holds. For every $t$ and every $\Pi$ for which
$(\beta,\Pi)$ satisfies Assumption~\ref{ass:ind}, a firm that is unmatched in
some outcome in $S^{B}(t,\Pi)$ earns zero in every outcome in $S^{B}(t,\Pi)$.
\end{theorem}

The proof of Theorem~\ref{thm:firm} uses Assumption~\ref{ass:sep} only through
Lemma~\ref{lem:opposition}. With Lemma~\ref{lem:oppcom} in its place it applies
without change.

Neither assumption can simply be dropped. Consider first
Assumption~\ref{ass:com}.
The assumption is an implication: if a type is relatively more eager to join a
firm, then it is at least as valuable to that firm. It fails when, for some
worker, some pair of firms and some pair of types, the premise holds and the
conclusion does not. The next proposition shows that such a failure is enough
to construct a violation of the Lone Wolf property.

Return to the environment of Proposition~\ref{prop:necessity}: one worker with
types $L$ and $H$ and a full-support prior, two uninformed firms $A$ and $B$, and
$\delta_s=a_A(s)-a_B(s)$. With a single worker, Assumption~\ref{ass:ind} holds
trivially. Label the firms and the types so that the failure of
Assumption~\ref{ass:com} occurs at firm $A$ and type $H$ is the more eager type.
The premise is then $\delta_H>\delta_L$, and the conclusion that fails is
$b_A(H)\ge b_A(L)$.

\begin{proposition}[Failure without \com]\label{prop:necessitycom}
In the environment of Proposition~\ref{prop:necessity}, suppose that
Assumption~\ref{ass:com} fails:
\[
\delta_H>\delta_L\qquad\text{and}\qquad b_A(H)<b_A(L).
\]
Then there are worker values with $a_A(s)-a_B(s)=\delta_s$ for $s\in\{L,H\}$ and
a nonempty open set of values $(b_B(L),b_B(H))\in\R^2$ of firm $B$ for which, at
each $t\in\{L,H\}$, the stratum $S^{B}(t,\Pi^{0})$ contains an outcome in which the
worker is matched with $A$ and firm $A$ earns a positive payoff, and an outcome
in which the worker is matched with $B$. For these values, firm-side payoff
Lone Wolf and matched-set Lone Wolf fail on $S^{B}(t,\Pi^{0})$.
\end{proposition}

Consider next Assumption~\ref{ass:ind}. In the following example every worker
satisfies Assumption~\ref{ass:com} in one of its two leading forms, and
firm-side payoff Lone Wolf fails because types are correlated.

\begin{example}[Correlated types]\label{ex:corr}
There are three workers, $1$, $2$ and $3$, and four firms, $A$, $B$, $C$ and
$D$. All firms are uninformed. Worker $2$ has two equally likely types, $L$ and
$H$. Workers $1$ and $3$ have two types each, $l$ and $h$, which are
independent conditional on worker $2$'s type. Worker $1$ has type $h$ with
probability $\tfrac34$ if worker $2$ has type $L$, and with probability
$\tfrac14$ if worker $2$ has type $H$. Worker $3$ has type $h$ with probability
$\tfrac14$ if worker $2$ has type $L$, and with probability $\tfrac34$ if
worker $2$ has type $H$. The prior has full support. The matching values are
\[
\begin{array}{c|c|c}
\text{pair } (i,j) & a_{ij} & b_{ij}\\ \hline
(1,A) & 10 & 2\\
(1,B) & 10 & 0 \text{ at } l,\ \ 40 \text{ at } h\\
(2,B) & 10 & 26\\
(2,C) & 10 \text{ at } L,\ \ 20 \text{ at } H & 22\\
(3,C) & 10 & 0 \text{ at } l,\ \ 40 \text{ at } h\\
(3,D) & 10 & 1
\end{array}
\]
and $a_{ij}=b_{ij}=-100$, at every type, for every other pair. Thus
$\E[b_{1B}]=\E[b_{3C}]=20$. Conditional on worker $2$ having type $L$, firm $B$
expects $30$ from worker $1$ and firm $C$ expects $10$ from worker $3$;
conditional on type $H$, firm $B$ expects $10$ from worker $1$ and firm $C$
expects $30$ from worker $3$.

The values of workers $1$ and $3$ do not depend on their types, so for them the
premise of Assumption~\ref{ass:com} never holds. Firms' values of worker $2$ do
not depend on his type, so for him its conclusion always holds. Hence
Assumption~\ref{ass:com} holds. Assumption~\ref{ass:ind} fails: worker $2$'s
type $L$ is good news about worker $1$, and his type $H$ is good news about
worker $3$.
\end{example}

\begin{proposition}[Failure without posterior independence]\label{prop:necessityind}
In Example~\ref{ex:corr}, for every $t\in T$ the stratum $S^{B}(t,\Pi^{0})$
contains the following two outcomes:
\begin{enumerate}[label=(\alph*)]
\item worker $1$ is matched with $A$ at the wage $0$, worker $2$ with $B$ at the
wage $5$, and worker $3$ with $C$ at the wage $2$, and firm $D$ is unmatched;
\item worker $1$ is matched with $B$ at the wage $3$, worker $2$ with $C$ at the
wage $0$, and worker $3$ with $D$ at the wage $0$, and firm $A$ is unmatched.
\end{enumerate}
Firm $A$ earns $2$ in (a) and is unmatched in (b). Hence firm-side payoff Lone
Wolf and matched-set Lone Wolf fail on $S^{B}(t,\Pi^{0})$, although
Assumption~\ref{ass:com} holds.
\end{proposition}

In outcome (a), firm $C$ would like to poach worker $2$. His eager type $H$
accepts a wage just above $-5$, and firm $C$ would earn almost $27$ from him,
well above its payoff of $18$. But only type $H$ accepts that wage, and type
$H$ is good news about worker $3$, whom firm $C$ would have to give up:
conditional on it, firm $C$'s current match is worth $28$. A wage that both
types accept is too expensive. Firm $B$ faces the same problem in outcome (b),
with type $L$ and worker $1$. If firms compared an offer with their payoffs
$v_j(t)$ instead of their re-evaluated payoffs, both outcomes would be
blocked. The mechanism is not adverse selection, because acceptance is not bad
news about the worker who moves. It is good news about the worker who would be
replaced.

For the generic result, fix $t$, $\Pi$, $\beta$ and $a$, and let $\calB$ denote
the set of firm values $b$ for which Assumption~\ref{ass:com} holds. It is a
polyhedral cone, and it is the whole space of firm values when $a$ is additively
separable.

\begin{theorem}[Generic Lone Wolf]\label{thm:genericcom}
Fix $t$, $\Pi$, $\beta$ and $a$, and suppose that Assumption~\ref{ass:ind}
holds. There is a finite union $\mathcal H$ of affine
hyperplanes in the space of firm values, none of which contains $\calB$, such
that for every $b\in\calB\setminus\mathcal H$ all outcomes in $S^{B}(t,\Pi)$ have
the same set of unmatched agents.
\end{theorem}

Since $\calB$ is a polyhedral cone, the exceptional firm values form a closed
subset of $\calB$ with empty relative interior and measure zero in the linear
span of $\calB$. With three or more firms $\calB$ can have lower dimension than
the space of firm values, so genericity is understood relative to $\calB$.

The lattice property requires one additional step. Let $(\mu,p)$ and
$(\hat\mu,\hat p)$ be outcomes in $S^{B}(t,\Pi)$ with the same set of unmatched
agents, and define their components as in Section~\ref{sec:lattice}. Under
additive separability, a worker's gain from one outcome to the other does not
depend on his type, and Lemma~\ref{lem:common} follows. Under
Assumption~\ref{ass:com} this is no longer an assumption, but it remains true
for the workers of a component.

\begin{lemma}[Constant differences]\label{lem:constant}
Suppose Assumptions~\ref{ass:com} and~\ref{ass:ind} hold, and let $(\mu,p)$ and $(\hat\mu,\hat p)$
be outcomes in $S^{B}(t,\Pi)$ with the same set of unmatched agents. For every
component $C$ there is a number $d_C$ such that, for every $t'\in\cK(t)$,
\begin{align*}
\hat u_i(t')-u_i(t')&=d_C\quad\text{for every worker } i\in C,\\
v_j(t')-\hat v_j(t')&=d_C\quad\text{for every firm } j\in C .
\end{align*}
\end{lemma}

The first step of the proof is the cycle of
inequalities from the proof of Lemma~\ref{lem:common}, which shows that at each
type profile the gains are equal on a component. The new step shows that the
common gain cannot vary with the type profile. Across type profiles, Assumption~\ref{ass:com} makes a firm's loss move weakly
against the gains of the two workers it employs in the two outcomes, so the
common value can only be constant.

\begin{theorem}[Lattice]\label{thm:latticecom}
Suppose Assumptions~\ref{ass:com} and~\ref{ass:ind} hold.
\begin{enumerate}[label=(\roman*)]
\item If two outcomes in $S^{B}(t,\Pi)$ have the same set of unmatched agents,
then they have a join and a meet.
\item If matched-set Lone Wolf holds on $S^{B}(t,\Pi)$, then the lattice
property holds on $S^{B}(t,\Pi)$, and a nonempty $S^{B}(t,\Pi)$ contains a
worker-optimal and a firm-optimal outcome. By Theorem~\ref{thm:genericcom}, this
is the case for all $b\in\calB\setminus\mathcal H$.
\end{enumerate}
\end{theorem}

The join and the meet are built as
in Proposition~\ref{prop:lattice}, by recombining the two outcomes component by
component. A blocking offer to an employed worker may now screen his types, so
Lemma~\ref{lem:noscreen} is no longer available. On the firm's side,
Lemma~\ref{lem:noincumbent} replaces it: the firm's re-evaluated payoff on the
willingness set is its payoff. With this, the argument for the meet is unchanged.
For the join, Lemma~\ref{lem:constant} takes care of the worker's side:
lowering the offer by the worker's constant gain leaves its willingness set
unchanged, so the block can be transferred to one of the original outcomes.

\section{Markets without transfers}\label{sec:ntu}

The questions of this paper are also studied in a model without transfers by
\citet{hu2026structure}. This section compares the two models. The mechanism
behind the failures is the same, and so is the kind of condition that removes
it. The conditions themselves differ: without transfers they concern workers'
\emph{rankings} of firms, and with transfers they concern the \emph{differences}
in workers' values. The conclusions differ as well, in ways that can be traced
to transfers.

I follow \citet{hu2026structure}, whose notation Section~\ref{sec:model} was
chosen to match. The primitives are the same: workers $I$, firms $J$, type
profiles $T$ with the full-support prior $\beta$, and values $a_{ij}(t_i)$ and
$b_{ij}(t_i)$, with the value of being unmatched normalized to zero,
$a_{ii}\equiv0$ and $b_{jj}\equiv0$. There are no wages. A \emph{state}
$(\mu,t,\Pi)$ consists of a matching, a type profile and an information
structure, and an agent's payoff is the value of his match: $a_{i\mu(i)}(t)$ for
worker $i$, and $\E[b_{\mu(j)j}\mid\Pi_j(t)]$ for firm $j$.

\begin{definition}[\citealp{hu2026structure}]\label{def:ntu}
Fix a state $(\mu,t,\Pi)$.
\begin{enumerate}[label=(\roman*)]
\item The state is \textbf{individually rational} if $a_{i\mu(i)}(t)\ge0$ for every
worker $i$ and $\E[b_{\mu(j)j}\mid\Pi_j(t)]\ge0$ for every firm $j$.
\item For a worker $i$ and a firm $j\ne\mu(i)$, the \textbf{willingness set} is
\[
D_{\mu,ij}:=\bigl\{t'\in T:\ a_{ij}(t')>a_{i\mu(i)}(t')\bigr\}.
\]
The state is \textbf{blocked} by $(i,j)$ if
\[
a_{ij}(t)>a_{i\mu(i)}(t)
\qquad\text{and}\qquad
\E\bigl[b_{ij}\mid\Pi_j(t)\cap D_{\mu,ij}\bigr]>\E\bigl[b_{\mu(j)j}\mid\Pi_j(t)\cap D_{\mu,ij}\bigr].
\]
\item The state is \textbf{Bayesian stable} if $(\mu,t',\Pi)$ is individually
rational and not blocked for every $t'\in\cK(t)$.
\end{enumerate}
\end{definition}

As in Definition~\ref{def:blocked}, a firm that considers a new worker
re-evaluates its current worker on the willingness set. Part (iii) is the
common-knowledge form of his Definition~4; the two are equivalent by his
Proposition~1. As in \citet{hu2026structure}, preferences are \emph{strict}:
$a_{ij}(t)\neq a_{ij'}(t)$ for every worker $i$, every $t\in T$ and all distinct
$j,j'\in J\cup\{i\}$, and $\E[b_{ij}\mid D]\neq\E[b_{i'j}\mid D]$ for every firm
$j$, all distinct $i,i'\in I\cup\{j\}$ and every nonempty $D\subseteq T$.

Three further conditions are relevant.

\begin{definition}\label{def:conditions}
\begin{enumerate}[label=(\roman*)]
\item Workers' values are \textbf{multiplicatively separable} if for every worker
$i$ there are a function $g_i:T_i\to(0,\infty)$ and numbers $(h_{ij})_{j\in J}$
such that $a_{ij}(t_i)=g_i(t_i)\,h_{ij}$ for all $j$ and $t_i$.
\item Workers have \textbf{type-independent rankings} if, for every worker $i$,
all firms $j,\ell$ and all types $s,s'\in T_i$, $a_{ij}(s)>a_{i\ell}(s)$ if and only
if $a_{ij}(s')>a_{i\ell}(s')$.
\item Values satisfy \textbf{ordinal \com} if, for every worker $i$, all firms
$j\ne\ell$ and all types $s,s'\in T_i$,
\[
a_{ij}(s)>a_{i\ell}(s)\ \text{ and }\ a_{ij}(s')\le a_{i\ell}(s')
\quad\text{imply}\quad
b_{ij}(s)\ge b_{ij}(s').
\]
\end{enumerate}
\end{definition}

Type-independent rankings is the ordinal counterpart of additive separability:
it restricts the sign of $a_{ij}(t_i)-a_{i\ell}(t_i)$, where
Assumption~\ref{ass:sep} restricts its size. Ordinal \com{} is the ordinal
counterpart of Assumption~\ref{ass:com}: a type who prefers $j$ to $\ell$ is at
least as valuable to $j$ as a type who does not. Multiplicative separability is
a cardinal condition that is also cited in \cite{hu2026structure}.

\begin{proposition}\label{prop:relations}
The following implications hold:
\begin{enumerate}[label=(\roman*)]
\item Additive separability implies \com{} and type-independent rankings.
\item Multiplicative separability implies type-independent rankings.
\item \Com{} implies ordinal \com.
\item Type-independent rankings imply ordinal \com.
\end{enumerate}
No other implication holds among the five conditions, apart from those that
follow from (i)--(iv) by transitivity.
\end{proposition}

\begin{figure}[t]
\centering
\begin{tikzpicture}[
  box/.style={draw, rounded corners=2pt, align=center, inner sep=5pt,
              text width=3.5cm, minimum height=1.25cm, font=\small},
  imp/.style={-{Latex[length=2.4mm]}, line width=0.8pt, shorten >=2pt, shorten <=2pt},
  frame/.style={draw, dashed, rounded corners=4pt, inner sep=9pt}]
\node[box] (AS)  at (0,0)     {Additive separability\\ $a_{ij}=g_i(t_i)+h_{ij}$};
\node[box] (TIR) at (5.3,0)   {Type-independent rankings};
\node[box] (MS)  at (10.6,0)  {Multiplicative separability\\ $a_{ij}=g_i(t_i)\,h_{ij}$};
\node[box] (COM) at (0,-2.7)  {\Com};
\node[box] (OC)  at (5.3,-2.7){Ordinal\\ \com};
\draw[imp] (AS.east) -- (TIR.west);
\draw[imp] (MS.west) -- (TIR.east);
\draw[imp] (AS.south) -- (COM.north);
\draw[imp] (TIR.south) -- (OC.north);
\draw[imp] (COM.east) -- (OC.west);
\node[frame, fit=(AS)(COM)] (FL) {};
\node[frame, fit=(MS)(TIR)(OC)] (FR) {};
\node[font=\small\itshape, above=2pt of FL] {with transfers};
\node[font=\small\itshape, above=2pt of FR] {without transfers};
\end{tikzpicture}
\caption{Relations among the conditions. }\label{fig:conditions}
\end{figure}
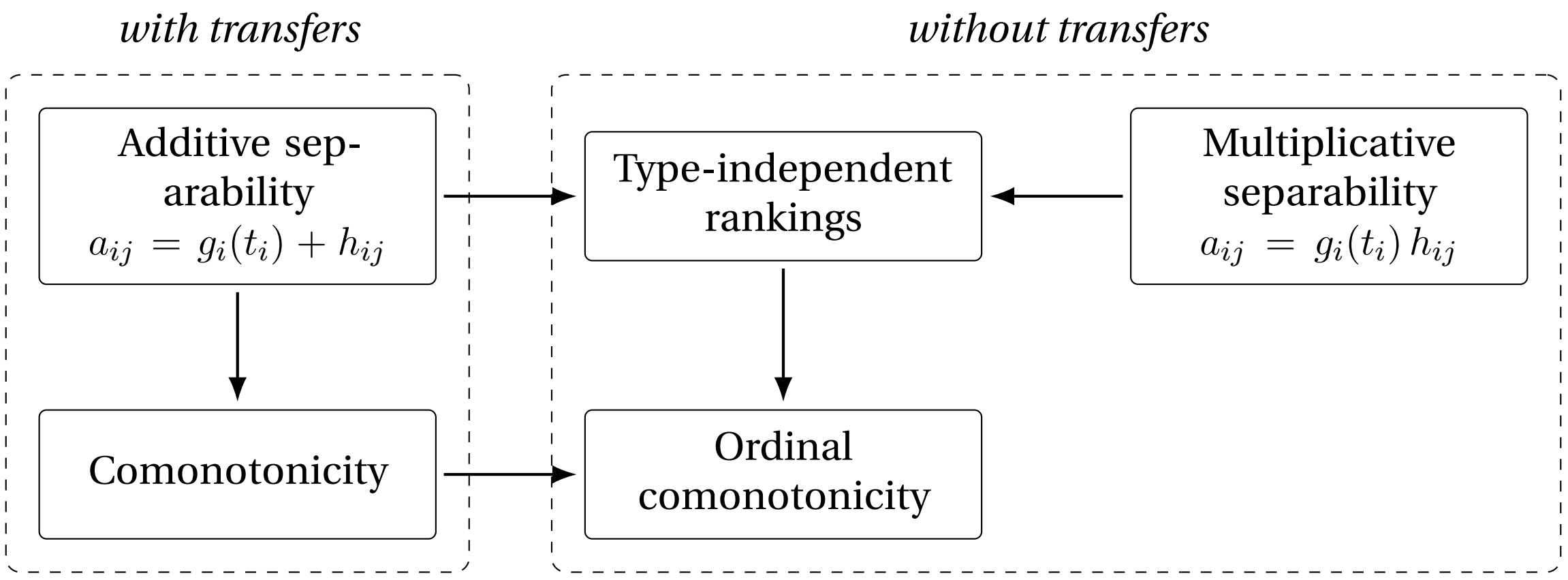

Figure~\ref{fig:conditions} summarizes the proposition.

Under complete information the two models are mirror images: the Lone Wolf and
lattice properties hold in both. One might therefore expect that whenever a
cardinal condition delivers the structural properties with transfers, its
ordinal counterpart delivers them without transfers. This is only partly true.

To state the lattice property without transfers, let $\mu$ and $\hat\mu$ be
matchings in $S^{B}(t,\Pi)$. Following \citet[Definition~5]{hu2026structure},
the \emph{worker join} $\lambda=\mu\vee_I\hat\mu$ is the function on $I\cup J$
that assigns to each worker $i$ the partner he prefers at $t$ among $\mu(i)$
and $\hat\mu(i)$, and to each firm $j$ the partner with the lower expected
value $\E[b_{\cdot j}\mid\Pi_j(t)]$ among $\mu(j)$ and $\hat\mu(j)$. The
\emph{worker meet} $\nu=\mu\wedge_I\hat\mu$ assigns to each worker his less
preferred partner and to each firm the partner with the higher expected value.
In general $\lambda$ and $\nu$ need not be matchings. The \emph{lattice
property} holds on $S^{B}(t,\Pi)$ if $\mu\vee_I\hat\mu$ and
$\mu\wedge_I\hat\mu$ are matchings in $S^{B}(t,\Pi)$ for all
$\mu,\hat\mu\in S^{B}(t,\Pi)$.

\begin{proposition}[Without transfers]\label{prop:ntu}
Consider the model without transfers.
\begin{enumerate}[label=(\roman*)]
\item If workers have type-independent rankings, then for every $t$ and every
$\Pi$ the set of unmatched agents is the same in all matchings in
$S^{B}(t,\Pi)$. The same conclusion holds under ordinal \com{} whenever
$(\beta,\Pi)$ satisfies Assumption~\ref{ass:ind}.
\item If workers have type-independent rankings, then the lattice property
holds on $S^{B}(t,\Pi)$ for every $t$ and every $\Pi$.
\item Under ordinal \com{} and Assumption~\ref{ass:ind} the lattice property
can fail: there are a market and matchings $\mu,\hat\mu\in S^{B}(t,\Pi)$ such
that $\mu\vee_I\hat\mu$ is a matching and does not belong to $S^{B}(t,\Pi)$.
\end{enumerate}
\end{proposition}

Part (i) gives two conditions under which the conclusion of Theorem~1 of
\citet{hu2026structure} holds; by Remark~\ref{rem:ntu} it can fail when neither
holds and firms do not observe the types of their own workers. Part (ii) is
closely related to Theorems~2 and~$2'$ of \citet{hu2026structure}: additive and
multiplicative separability both imply type-independent rankings
(Proposition~\ref{prop:relations}), and the present proof makes explicit that
this type-independent ranking property is sufficient for the recombination
argument.

Thus the ordinal counterpart of additive separability delivers both properties
without transfers. The ordinal counterpart of \com{} delivers the Lone Wolf
property but not the lattice property, although with transfers \com{} and
posterior independence deliver the lattice property whenever all outcomes match the same
agents (Theorem~\ref{thm:latticecom}), as the matchings in part (iii) do.
Posterior independence does not explain the difference. It is imposed with
\com{} in both models, and for the same reason: in both, a deviating firm re-evaluates
its current worker on the event that the new worker is willingness to move.

\section{Concluding remarks}\label{sec:conclusion}

With transfers and incomplete information, a wage offer selects the types who
accept it. This paper has traced the consequences for two structural properties
of stable outcomes. In general, a firm can be unmatched in one stable outcome
and earn a positive payoff in another, and stable payoffs need not form a
lattice, even within a common-information stratum. Under additive separability,
firm-side payoff Lone Wolf holds on every stratum. Matched-set Lone Wolf and
the lattice property hold generically in the relevant space of firm values,
outside a finite union of affine hyperplanes. On the exceptional set, changes
in unmatched sets are constrained by binding alternating paths, and a worker
who moves between employment and unemployment earns a precisely characterized
information rent. The firm-side theorem, the generic Lone Wolf theorem, and the
lattice theorem extend to comonotonicity of relative preferences when posterior
beliefs are independent across workers.

The main lesson is that the classical properties rest on two informational
conditions that complete information hides. A deviation must not be bad news
about the worker who is willingness to move, nor good news about the incumbent who
would be replaced. Comonotonicity addresses the first concern: acceptance is
never bad news about the moving worker. Posterior independence across workers
addresses the second: conditioning on one worker's acceptance does not change
the posterior marginal beliefs about the other workers. Under complete
information both steps are automatic. With transfers, the first condition must
be stated in terms of the intensity of workers' preferences, because wages can
screen on intensity where rankings alone cannot.

Several questions are open.

\paragraph{Observability within matched pairs.} I have not assumed that a firm
learns the type of its own worker. Suppose that it does, in the sense of
\citet[Remark~1]{hu2026structure}: in every state, the cell $\Pi_j(t')$ of a
matched firm $j$ contains only type profiles at which the type of its worker is
the same as at $t'$. Two outcomes with different matchings then belong to the
same stratum only if every firm knows the types of both of the workers it
employs in them. In that case a firm that approaches a worker whom it employs
in the other outcome already knows his type, the offer in the proof of
Lemma~\ref{lem:opposition} is accepted on the whole of the firm's cell, and
pairwise opposition holds without any restriction on values. The proofs of
Theorems~\ref{thm:firm} and~\ref{thm:generic} then apply as they stand. The
failures of Section~\ref{sec:failure} thus rely on firms being uncertain about
the workers they employ, as in \citep{liu2020stability}; Example~\ref{ex:failure}
does not survive observability, which forces both firms to be fully informed.
Whether the lattice results hold under observability without a restriction on
values is open.

\paragraph{Richer markets.} The analysis is for one-to-one matching with private
information on one side. In many-to-one matching a firm's inference from one
worker's willingness to move interacts with its other workers: the
re-evaluation of the current partner in the blocking condition would extend to
all of the firm's current workers. With private
information on both sides, a worker also learns from a firm's offer, and the
asymmetry between the firm-side and worker-side results may change. The
stochastic matchings of \citet{liu2020stability}, in which firms hold private
signals, are a natural intermediate case.

\paragraph{Correlated types.} Under \com{} the results use posterior
independence across workers, and Example~\ref{ex:corr} shows that the firm-side
theorem can fail without it. Which forms of correlation preserve the lattice results is
open.

\paragraph{Existence for a given information structure.} The results describe a set
$S^{B}(t,\Pi)$ when it is nonempty. Bayesian stable outcomes always exist for
some information structure, but a given $\Pi$ may support none. A description
of the information structures that support stable outcomes, and of how the
structural properties vary across them, would complement the results here.

\appendix
\titleformat{\section}{\centering\normalfont\normalsize\scshape}{Appendix \thesection:}{0.5em}{}
\renewcommand{\thelemma}{A.\arabic{lemma}}
\renewcommand{\thecorollary}{A.\arabic{corollary}}
\renewcommand{\theHlemma}{A.\arabic{lemma}}
\renewcommand{\theHcorollary}{A.\arabic{corollary}}
\setcounter{lemma}{0}
\setcounter{corollary}{0}
\section{Proofs}\label{app:proofs}

The proofs appear in the order of the results.

\subsection{Proofs for Section~\ref{sec:failure}}

\begin{proof}[Proof of Proposition~\ref{prop:CI}]
Let $(\mu,p)$ and $(\hat\mu,\hat p)$ be outcomes in $S^{B}(t,\Pi^{CI})$, with
payoffs $(u,v)$ and $(\hat u,\hat v)$ at $t$. Let
$V(\mu):=\sum_{i:\,\mu(i)\in J}\alpha_{i\mu(i)}$ be the total surplus of $\mu$,
and let $U(\mu)$ be the set of agents who are unmatched under $\mu$. Since
matched partners share their surplus and unmatched agents earn zero,
$\sum_i u_i+\sum_j v_j=V(\mu)$ and $\sum_i\hat u_i+\sum_j\hat v_j=V(\hat\mu)$.
Group the payoffs $(\hat u,\hat v)$ according to $\mu$. No blocking in
$(\hat\mu,\hat p)$ gives $\hat u_i+\hat v_{\mu(i)}\ge\alpha_{i\mu(i)}$ for every
worker $i$ who is matched under $\mu$. Hence
\[
V(\hat\mu)\ \ge\ V(\mu)+\sum_{i\in U(\mu)\cap I}\hat u_i+\sum_{j\in U(\mu)\cap J}\hat v_j .
\]
Exchanging the roles of the two outcomes,
\[
V(\mu)\ \ge\ V(\hat\mu)+\sum_{i\in U(\hat\mu)\cap I}u_i+\sum_{j\in U(\hat\mu)\cap J}v_j .
\]
Adding the two inequalities, the four sums add up to at most zero. Each term is
nonnegative by individual rationality, so each term is zero. Thus an agent who
is unmatched under $\mu$ earns zero in $(\hat\mu,\hat p)$.
\end{proof}

\begin{proof}[Proof of Proposition~\ref{prop:failure}]
In every case below the firm that makes an offer is unmatched, so its
re-evaluated payoff is zero by Lemma~\ref{lem:reeval}(i).

\emph{The worker is unmatched.} Firm $A$ can offer the wage $q=-5$. Both types
accept, since $6-5>0$ and $8-5>0$, so $D_{A}(q)=T$ and firm $A$ expects
$\tfrac52+5>0$. The state is blocked.

\emph{The worker is matched with $A$ at the wage $p$.} Individual rationality at
both types requires $6+p\ge0$ and $\tfrac52-p\ge0$, i.e., $-6\le p\le\tfrac52$. The
only candidate blocking pair is the worker and firm $B$. Type $t^0$ accepts an
offer $q$ from $B$ if and only if $8+q>6+p$, i.e., $q>p-2$. Type $t^1$ accepts if
and only if $1+q>8+p$, i.e., $q>p+7$. There are three cases.
\begin{itemize}
\item If $q\le p-2$, no type accepts.
\item If $p-2<q\le p+7$, then $D_{B}(q)=\{t^0\}$. The offer can block only at
$t^0$, and it does so if and only if $b_B(t^0)-q=-1-q>0$. Such a $q$ exists if and
only if $p-2<-1$, i.e., $p<1$.
\item If $q>p+7$, then $D_{B}(q)=T$, and the offer blocks if and only if
$\tfrac92-q>0$. Such a $q$ exists if and only if $p<-\tfrac52$.
\end{itemize}
Hence the state is IR and not blocked at both types if and only if
$1\le p\le\tfrac52$.

\emph{The worker is matched with $B$ at the wage $p$.} Individual rationality
requires $1+p\ge0$ and $\tfrac92-p\ge0$. Type $t^1$ accepts an offer $q$ from $A$
if and only if $8+q>1+p$, i.e., $q>p-7$. Type $t^0$ accepts if and only if
$6+q>8+p$, i.e., $q>p+2$.
\begin{itemize}
\item If $q\le p-7$, no type accepts.
\item If $p-7<q\le p+2$, then $D_{A}(q)=\{t^1\}$. The offer blocks at $t^1$ if and
only if $b_A(t^1)-q=-3-q>0$. Such a $q$ exists if and only if $p<4$.
\item If $q>p+2$, then $D_{A}(q)=T$, and the offer blocks if and only if
$\tfrac52-q>0$. Such a $q$ exists if and only if $p<\tfrac12$.
\end{itemize}
Hence the state is IR and not blocked at both types if and only if
$4\le p\le\tfrac92$.
\end{proof}

\begin{proof}[Proof of Corollary~\ref{cor:failure}]
Part (i) is immediate from Proposition~\ref{prop:failure}. For part (ii), write
payoff vectors as $(u;v_A,v_B)$. At $t^1$ the two outcomes have payoffs
$(9;\tfrac32,0)$ and $(5;0,\tfrac12)$, so a join would have payoffs $(9;0,0)$ and a
meet $(5;\tfrac32,\tfrac12)$. At $t^0$ they have payoffs $(7;\tfrac32,0)$ and
$(12;0,\tfrac12)$, so a join would have payoffs $(12;0,0)$ and a meet
$(7;\tfrac32,\tfrac12)$. No outcome in the stratum gives both firms a positive
payoff, which rules out the meets. In the outcomes in (a), firm $A$ earns zero
only at the wage $\tfrac52$, where the worker earns $\tfrac{21}{2}$ at $t^1$ and
$\tfrac{17}{2}$ at $t^0$. In the outcomes in (b), firm $B$ earns zero only at the
wage $\tfrac92$, where the worker earns $\tfrac{11}{2}$ at $t^1$ and $\tfrac{25}{2}$
at $t^0$. This rules out the joins.
\end{proof}

\subsection{Proofs for Section~\ref{sec:separable}}

Throughout this part, as in Section~\ref{sec:firms}, $t$ and $\Pi$ are fixed,
$(\mu,p)$ and $(\hat\mu,\hat p)$ are outcomes in $S^{B}(t,\Pi)$ with payoffs
$(u,v)$ and $(\hat u,\hat v)$, and $\hat D_{ij}(q)$ and
$\hat v_j(\cdot\mid\cdot)$ are the willingness sets and the re-evaluated payoffs in
the state $(\hat\mu,\hat p,t,\Pi)$.

\begin{proof}[Proof of Lemma~\ref{lem:rehiring}]
Suppose that $\hat v_j(t)<v_j(t)$. Since $(\mu,p,t,\Pi)$ is Bayesian stable,
$a_{ij}(t')+p_i\ge0$ for every $t'\in\cK(t)$. Consider the state
$(\hat\mu,\hat p,t,\Pi)$ and the offer $q=p_i+\varepsilon$ from $j$ to $i$, where
$\varepsilon>0$. For every $t'\in\cK(t)$,
$a_{ij}(t')+q>0=\hat u_i(t')$. Hence $\cK(t)\subseteq\hat D_{ij}(q)$. Since
$\Pi_j(t)\subseteq\cK(t)$, firm $j$'s re-evaluated payoff on $\hat D_{ij}(q)$ is
$\hat v_j(t)$ by Lemma~\ref{lem:reeval}(ii), and
\[
\E\bigl[b_{ij}\mid\Pi_j(t)\cap\hat D_{ij}(q)\bigr]-q
=\E\bigl[b_{ij}\mid\Pi_j(t)\bigr]-p_i-\varepsilon
=v_j(t)-\varepsilon .
\]
For $\varepsilon$ small enough, $v_j(t)-\varepsilon>\hat v_j(t)$. Thus $(i,j)$
blocks $(\hat\mu,\hat p,t,\Pi)$, a contradiction.
\end{proof}

\begin{proof}[Proof of Lemma~\ref{lem:opposition}]
Suppose that $\hat u_i(t)+\hat v_j(t)<S_{ij}(t)$. Consider the state
$(\hat\mu,\hat p,t,\Pi)$ and the offer
$q_\varepsilon:=\hat u_i(t)-a_{ij}(t)+\varepsilon$ from $j$ to $i$, where
$\varepsilon>0$. Since $\hat u_i(t)=a_{i\ell}(t)+\hat p_i$, Assumption~\ref{ass:sep}
gives $q_\varepsilon=\hat p_i+h_{i\ell}-h_{ij}+\varepsilon$. By
Lemma~\ref{lem:noscreen}, $\hat D_{ij}(q_\varepsilon)=T$, so firm $j$'s
re-evaluated payoff is $\hat v_j(t)$ by Lemma~\ref{lem:reeval}(ii), and
\[
\E\bigl[b_{ij}\mid\Pi_j(t)\cap\hat D_{ij}(q_\varepsilon)\bigr]-q_\varepsilon
=\E\bigl[b_{ij}\mid\Pi_j(t)\bigr]+a_{ij}(t)-\hat u_i(t)-\varepsilon
=S_{ij}(t)-\hat u_i(t)-\varepsilon .
\]
For $\varepsilon$ small enough the right-hand side exceeds $\hat v_j(t)$. The
worker strictly gains at $t$, since $a_{ij}(t)+q_\varepsilon=\hat u_i(t)+\varepsilon$.
Thus $(i,j)$ blocks $(\hat\mu,\hat p,t,\Pi)$, a contradiction.
\end{proof}

\begin{proof}[Proof of Theorem~\ref{thm:firm}]
Suppose not. Then there are outcomes $(\mu,p)$ and $(\hat\mu,\hat p)$ in
$S^{B}(t,\Pi)$ and a firm $j_0$ that is unmatched under $\hat\mu$ and has
$\gamma:=v_{j_0}(t)>0$. For each worker $i$ and each firm $j$, write
\[
\Delta u_i:=\hat u_i(t)-u_i(t)
\qquad\text{and}\qquad
\Delta v_j:=v_j(t)-\hat v_j(t).
\]
Note that $\Delta v_{j_0}=\gamma$.

Consider the alternating path of $\mu$ and $\hat\mu$ that starts at $j_0$:
\[
j_0,\quad i_1:=\mu(j_0),\quad j_1:=\hat\mu(i_1),\quad i_2:=\mu(j_1),\quad j_2:=\hat\mu(i_2),\ \dots
\]
Each agent has at most one partner under $\mu$ and at most one under $\hat\mu$,
and $j_0$ has no partner under $\hat\mu$. Hence the agents on the path are
distinct for as long as the path continues. I show by induction on $r\ge1$ that
$i_r$ and $j_r$ are well defined, that $i_r\in I$ and $j_r\in J$, and that
\[
\Delta u_{i_r}\ge\gamma
\qquad\text{and}\qquad
\Delta v_{j_r}\ge\gamma .
\]
This contradicts the finiteness of $I\cup J$.

Let $r\ge1$ and suppose that $j_{r-1}\in J$ and $\Delta v_{j_{r-1}}\ge\gamma$; this
holds for $r=1$.

\emph{Firm $j_{r-1}$ is matched under $\mu$.} Indeed,
$v_{j_{r-1}}(t)\ge\gamma+\hat v_{j_{r-1}}(t)\ge\gamma>0$. So
$i_r=\mu(j_{r-1})\in I$.

\emph{Worker $i_r$ is matched under $\hat\mu$.} Otherwise
Lemma~\ref{lem:rehiring} gives $\hat v_{j_{r-1}}(t)\ge v_{j_{r-1}}(t)$, which
contradicts $\Delta v_{j_{r-1}}>0$. So $j_r=\hat\mu(i_r)\in J$, and
$j_r\ne j_{r-1}$ because the agents on the path are distinct.

\emph{$\Delta u_{i_r}\ge\gamma$.} Lemma~\ref{lem:opposition}, applied to the pair
$(i_r,j_{r-1})$, which is matched under $\mu$, gives
$\hat u_{i_r}(t)+\hat v_{j_{r-1}}(t)\ge u_{i_r}(t)+v_{j_{r-1}}(t)$, that is,
$\Delta u_{i_r}\ge\Delta v_{j_{r-1}}\ge\gamma$.

\emph{$\Delta v_{j_r}\ge\gamma$.} Lemma~\ref{lem:opposition} with the roles of the
two outcomes exchanged, applied to the pair $(i_r,j_r)$, which is matched under
$\hat\mu$, gives $u_{i_r}(t)+v_{j_r}(t)\ge\hat u_{i_r}(t)+\hat v_{j_r}(t)$, that is,
$\Delta v_{j_r}\ge\Delta u_{i_r}\ge\gamma$.
\end{proof}

\begin{proof}[Proof of Proposition~\ref{prop:necessity}]
Choose wages $p_A$ and $p_B$ such that
\[
\delta_L<p_B-p_A<\delta_H,
\]
and large enough that $a_A(s)+p_A>0$ and $a_B(s)+p_B>0$ for both types $s$. Then
$p_A<p_B-\delta_L$ and $p_B<p_A+\delta_H$, so there are firm values that satisfy
\begin{align*}
p_A&<\E[b_A]<p_B-\delta_L, & b_A(H)&<p_B-\delta_H,\\
p_B&<\E[b_B]<p_A+\delta_H, & b_B(L)&<p_A+\delta_L;
\end{align*}
since the prior has full support, one can first fix $b_A(H)$ and $b_B(L)$ and then
choose $b_A(L)$ and $b_B(H)$ to obtain the required expectations. I show that,
for such firm values, the outcome in which the worker is matched with $A$ at the
wage $p_A$ and the outcome in which he is matched with $B$ at the wage $p_B$ are
both in the stratum. Both firms are uninformed, so each state must be
individually rational and not blocked at both types.

\emph{The worker is matched with $A$ at the wage $p_A$.} The worker's payoff is
positive at both types, and firm $A$ earns $\E[b_A]-p_A>0$. Type $s$ accepts an
offer $q$ from $B$ if and only if $q>p_A+\delta_s$, so type $L$ accepts first. An
offer accepted by type $L$ alone gives $B$ less than $b_B(L)-p_A-\delta_L<0$. An
offer accepted by both types satisfies $q>p_A+\delta_H$ and gives $B$ less than
$\E[b_B]-p_A-\delta_H<0$. So firm $B$, which is unmatched, does not block.

\emph{The worker is matched with $B$ at the wage $p_B$.} The worker's payoff is
positive at both types, and firm $B$ earns $\E[b_B]-p_B>0$. Type $s$ accepts an
offer $q$ from $A$ if and only if $q>p_B-\delta_s$, so type $H$ accepts first. An
offer accepted by type $H$ alone gives $A$ less than $b_A(H)-p_B+\delta_H<0$. An
offer accepted by both types satisfies $q>p_B-\delta_L$ and gives $A$ less than
$\E[b_A]-p_B+\delta_L<0$. So firm $A$, which is unmatched, does not block.

All the inequalities imposed on $b$ are strict, so they continue to hold on a
neighborhood of the chosen values, with the same wages.
\end{proof}

\begin{proof}[Proof of Proposition~\ref{prop:worker}]
Both firms are uninformed, so each state must be IR and not blocked at both
types of worker $1$. Worker $2$ has a single type. Hence firm $A$'s value of
its worker, who is worker $2$ in both outcomes, is constant, and an offer to
worker $2$ is accepted at every type profile or at none. Firm $B$ is unmatched
in (b). By Lemma~\ref{lem:reeval}, every re-evaluated payoff in this proof
equals the corresponding payoff. In both outcomes worker $2$ earns $15+p_2\ge0$ and firm $A$
earns $5-p_2\ge0$. Worker $2$ accepts an offer $q$ from $B$ if and only if
$7+q>15+p_2$, and $B$ gains only if $1-q>0$; no such $q$ exists because
$p_2\ge-7$. This rules out the pair $(2,B)$ in both outcomes.

\emph{Outcome (b).} Worker $1$ of type $t^0$ accepts an offer $q$ if and only if
$q>1$, and type $t^1$ if and only if $q>3$. If $1<q\le3$ only $t^0$ accepts. Firm
$B$ would earn $-1-q<0$. Firm $A$ would earn $3-q<2\le5-p_2$, since $p_2\le3$.
If $q>3$ both types accept. Firm $B$ would earn $3-q<0$ and firm $A$ would earn
$\tfrac32-q<0\le5-p_2$. So neither firm blocks with worker $1$.

\emph{Outcome (a).} Worker $1$ earns $-1+3=2$ at $t^0$ and $-3+3=0$ at $t^1$, and
firm $B$ earns $3-3=0$, so the state is IR. Since $a_{1A}=a_{1B}$ at both types,
both types of worker $1$ accept an offer $q$ from $A$ if and only if $q>3$, and
then $A$ would earn $\tfrac32-q<0\le5-p_2$. So $(1,A)$ does not block.
\end{proof}

\begin{proof}[Proof of Lemma~\ref{lem:pinned}]
Let $m:=\max_{s\in\cK(t)}\bigl(-a_{ij}(s)\bigr)$. Individual rationality of
$(\mu,p,s,\Pi)$ for every $s\in\cK(t)$ gives $p_i\ge m$. Suppose that $p_i>m$,
and let $q\in(m,p_i)$. In the state $(\hat\mu,\hat p,t,\Pi)$, every
$s\in\cK(t)$ satisfies $a_{ij}(s)+q>a_{ij}(s)+m\ge0=\hat u_i(s)$, so
$\cK(t)\subseteq\hat D_{ij}(q)$ and
\[
\E\bigl[b_{ij}\mid\Pi_j(t)\cap\hat D_{ij}(q)\bigr]-q
=\E\bigl[b_{ij}\mid\Pi_j(t)\bigr]-q
>\E\bigl[b_{ij}\mid\Pi_j(t)\bigr]-p_i=v_j(t)=\hat v_j(t).
\]
By Lemma~\ref{lem:reeval}(ii), $\hat v_j(t)$ is also firm $j$'s re-evaluated
payoff on $\hat D_{ij}(q)$.
Thus $(i,j)$ blocks $(\hat\mu,\hat p,t,\Pi)$, a contradiction. Hence $p_i=m$ and
$u_i(t)=a_{ij}(t)+m=\rho_{ij}(t)$.
\end{proof}

\begin{proof}[Proof of Lemma~\ref{lem:binding}]
Write $\Delta u_i:=\hat u_i(t)-u_i(t)$ and $\Delta v_j:=v_j(t)-\hat v_j(t)$. The
agents on the path alternate between firms and workers, and its edges alternate
between $\mu$ and $\hat\mu$. Hence either every edge from a firm to the next
agent belongs to $\mu$ and every edge from a worker to the next agent belongs to
$\hat\mu$, or the same holds for the reversed sequence. Reversing a path leaves
$F_c$ unchanged if $L$ is odd and multiplies it by $-1$ if $L$ is even, so it
does not change $H_c$. I may therefore assume that $c$ is oriented in this way.
Then a worker on the path who is matched under both matchings is preceded by his
partner under $\mu$ and followed by his partner under $\hat\mu$. A first agent who
is a firm is unmatched under $\hat\mu$, a first agent who is a worker is
unmatched under $\mu$, a last agent who is a firm is unmatched under $\mu$, and a
last agent who is a worker is unmatched under $\hat\mu$. Moreover, $P(c)$ is
either the set of edges of $c$ that belong to $\mu$ or the set of those that
belong to $\hat\mu$, so it suffices to prove that
\begin{equation}\label{eq:path}
\sum_{e\in c\cap\mu}S_e(t)-\sum_{e\in c\cap\hat\mu}S_e(t)=\rho^{\mathrm{end}}-\rho^{\mathrm{start}},
\end{equation}
where $\rho^{\mathrm{end}}:=\rho_e(t)$ if the last agent is a worker with edge $e$,
and $\rho^{\mathrm{end}}:=0$ otherwise, and $\rho^{\mathrm{start}}$ is defined in the
same way for the first agent. Indeed, the left-hand side minus the right-hand
side equals $F_c(b)$ or $-F_c(b)$.

\emph{Step 1: payoff differences are nondecreasing along the path.} Let $i$ be a
worker on the path who is matched under both matchings, with $\mu(i)=j$ and
$\hat\mu(i)=\hat\jmath$. Lemma~\ref{lem:opposition} applied to $(i,j)$ gives
$\Delta u_i\ge\Delta v_j$. Applied to $(i,\hat\jmath)$ with the roles of the
outcomes exchanged, it gives $\Delta v_{\hat\jmath}\ge\Delta u_i$. Hence
$\Delta v_j\le\Delta u_i\le\Delta v_{\hat\jmath}$.

\emph{Step 2: the first firm on the path has $\Delta v\ge0$.} If the first agent
is a firm $j$, it is unmatched under $\hat\mu$, so $\Delta v_j=v_j(t)\ge0$. If the
first agent is a worker $i$, he is unmatched under $\mu$ and matched under
$\hat\mu$ with the next agent $j$; Lemma~\ref{lem:rehiring} with the roles
exchanged gives $v_j(t)\ge\hat v_j(t)$.

\emph{Step 3: the last firm on the path has $\Delta v\le0$.} If the last agent is
a firm $j$, it is unmatched under $\mu$, so $\Delta v_j=-\hat v_j(t)\le0$. If the
last agent is a worker $i$, he is unmatched under $\hat\mu$ and matched under
$\mu$ with the preceding agent $j$; Lemma~\ref{lem:rehiring} gives
$\hat v_j(t)\ge v_j(t)$.

\emph{Step 4: all differences vanish.} By Steps 1--3, $\Delta v_j=0$ for every
firm $j$ on the path and $\Delta u_i=0$ for every worker $i$ on the path who is
matched under both matchings. In particular, a firm endpoint earns zero in
both outcomes.

\emph{Step 5: worker endpoints earn their rents.} If the last agent is a worker
$i$, matched under $\mu$ with $j$, then $\hat v_j(t)=v_j(t)$ by Step 4, and
Lemma~\ref{lem:pinned} gives $u_i(t)=\rho_{ij}(t)$. Symmetrically, if the first
agent is a worker $i$, matched under $\hat\mu$ with $j$, then
$\hat u_i(t)=\rho_{ij}(t)$.

\emph{Step 6: summation.} Matched partners share their interim surplus, so
\[
\sum_{e\in c\cap\mu}S_e(t)-\sum_{e\in c\cap\hat\mu}S_e(t)
=\sum_{\{i,j\}\in c\cap\mu}\bigl(u_i(t)+v_j(t)\bigr)
-\sum_{\{i,j\}\in c\cap\hat\mu}\bigl(\hat u_i(t)+\hat v_j(t)\bigr).
\]
On the right-hand side, a firm that is not an endpoint contributes
$v_j(t)-\hat v_j(t)=0$, and a firm endpoint contributes zero by Step 4. A worker
who is matched under both matchings contributes $u_i(t)-\hat u_i(t)=0$. A last
agent who is a worker contributes $u_i(t)=\rho^{\mathrm{end}}$, and a first agent
who is a worker contributes $-\hat u_i(t)=-\rho^{\mathrm{start}}$. This
is~\eqref{eq:path}.
\end{proof}

\begin{proof}[Proof of Theorem~\ref{thm:generic}]
(i) Suppose that two outcomes in $S^{B}(t,\Pi)$ have different sets of
unmatched agents. Some agent is then matched under exactly one of the two
matchings, so the graph defined before Lemma~\ref{lem:binding} has a connected
component $c$ that is a path. By Lemma~\ref{lem:binding}, $F_c(b)=0$, so
$b\in\mathcal H$. If all outcomes have the same set of unmatched agents, an agent
who is unmatched in one of them earns zero in all of them.

(ii) Let $e_1=\{i,j\}$ be the first edge of $c$. The agents on a path are
distinct, so no other edge of $c$ is the pair $\{i,j\}$, and $R_c$ does not
depend on $b$. Hence the only term of $F_c$ that involves the values
$(b_{ij}(s))_{s\in T_i}$ is $\E[b_{ij}\mid\Pi_j(t)]$. Adding a constant
$\lambda$ to $b_{ij}(s)$ for every $s\in T_i$ changes $F_c$ by $\lambda$. So $F_c$
is a non-constant affine function and $H_c$ is an affine hyperplane. A finite
union of affine hyperplanes is closed, has empty interior, and has measure
zero.
\end{proof}

\begin{proof}[Proof of Corollary~\ref{cor:perturb}]
If worker-side payoff Lone Wolf fails at $b$, some worker is unmatched in one
outcome in $S^{B}(t,\Pi)$ and earns a nonzero payoff in another. He is matched in
the second outcome, so the two outcomes have different sets of unmatched agents,
and $b\in\mathcal H$ by Theorem~\ref{thm:generic}(i). The set $\mathcal H$ does
not depend on $b$: it is determined by $a$, $\beta$, $t$ and $\Pi$. By
Theorem~\ref{thm:generic}(ii), its complement is open and dense, and
$\mathcal H$ has measure zero. So every $\varepsilon$-ball around $b$ contains
firm values $b'\notin\mathcal H$, and at each of them
Theorem~\ref{thm:generic}(i) applies.
\end{proof}

\begin{proof}[Proof of Corollary~\ref{cor:rent}]
Worker $i$ is matched under exactly one of the two matchings, so he is an
endpoint of a path component of the graph defined before
Lemma~\ref{lem:binding}, and firm $j$ is the adjacent agent on that path. The
claim is the last part of that lemma.
\end{proof}

\begin{proof}[Proof of Proposition~\ref{prop:latticefail}]
Consider the outcome in which worker $1$ is employed at the wage $3$. Worker $1$
earns $2$ at type $t^0$ and $0$ at type $t^1$, and the firm earns
$\E[b_1]-3=1$, so the outcome is individually rational at every type profile.
The only possible block is by the firm and worker $2$. The two workers' types
are independent, so worker $2$'s willingness to accept an offer does not change
the firm's expected value of worker $1$, and the firm's re-evaluated payoff is
$1$ on every willingness set. Type $t^0$ of worker $2$
accepts an offer $q$ if and only if $q>1$, and type $t^1$ if and only if $q>3$.
An offer $q\in(1,3]$ attracts only type $t^0$ and gives the firm $-q<1$. An offer
$q>3$ attracts both types and gives the firm $4-q<1$. So there is no block at any
type profile, and the outcome belongs to $S^{B}(t,\Pi^{0})$. By symmetry, so does
the outcome in which worker $2$ is employed at the wage $3$.

In the first outcome the payoffs at $t$ are $2$ for worker $1$, $0$ for worker
$2$ and $1$ for the firm, and in the second the two workers' payoffs are
exchanged. A join would give each worker $2$. The firm employs at most one
worker, so in every outcome one worker earns $0$. A meet would give each worker
$0$ and the firm $1$. A worker who is employed at a wage that is individually
rational at type $t^1$ earns at least $2$ at type $t^0$, so both workers would be
unemployed and the firm would earn $0$.
\end{proof}

\begin{proof}[Proof of Lemma~\ref{lem:common}]
For a worker $i$ who is matched under both matchings,
Assumption~\ref{ass:sep} gives
\[
\hat u_i(s)-u_i(s)=h_{i\hat\mu(i)}+\hat p_i-h_{i\mu(i)}-p_i=:\delta_i
\qquad\text{for every } s\in T .
\]
Let $t'\in\cK(t)$. Since $\cK(t')=\cK(t)$, both outcomes belong to
$S^{B}(t',\Pi)$, so Lemma~\ref{lem:opposition} applies at $t'$.

If $C=\{i,j\}$ is a pair that is matched under both matchings, then
$v_j(t')-\hat v_j(t')=\hat p_i-p_i=\delta_i$, and I set $d_C:=\delta_i$.

Otherwise $C$ is a cycle $i_1,j_1,i_2,j_2,\dots,i_k,j_k$ with $k\ge2$,
$\mu(i_r)=j_r$ and $\hat\mu(i_r)=j_{r+1}$, where $j_{k+1}:=j_1$. Write
$\Delta v_j(t'):=v_j(t')-\hat v_j(t')$. Lemma~\ref{lem:opposition} at $t'$, applied
to the pair $(i_r,j_r)$, gives $\delta_{i_r}\ge\Delta v_{j_r}(t')$. Applied to the
pair $(i_r,j_{r+1})$ with the roles of the two outcomes exchanged, it gives
$\Delta v_{j_{r+1}}(t')\ge\delta_{i_r}$. Hence
\[
\Delta v_{j_1}(t')\le\delta_{i_1}\le\Delta v_{j_2}(t')\le\delta_{i_2}\le\dots\le\delta_{i_k}\le\Delta v_{j_1}(t'),
\]
and all these numbers are equal. Their common value $d_C$ does not depend on
$t'$, because the numbers $\delta_{i_r}$ do not.
\end{proof}

\begin{proof}[Proof of Proposition~\ref{prop:lattice}]
Let $(\mu,p)$ and $(\hat\mu,\hat p)$ be the two outcomes and let $d_C$ be as in
Lemma~\ref{lem:common}. Define the outcome $(\mu^\vee,p^\vee)$ as follows: on
each component $C$ with $d_C>0$ it coincides with $(\hat\mu,\hat p)$, on every
other component it coincides with $(\mu,p)$, and the agents who are unmatched
under $\mu$ are unmatched. Define $(\mu^\wedge,p^\wedge)$ in the same way with
$d_C<0$ in place of $d_C>0$. Both are well defined, because each matching
matches the agents of a component among themselves. Denote their payoffs by
$(u^\vee,v^\vee)$ and $(u^\wedge,v^\wedge)$. By Lemma~\ref{lem:common}, for every
$t'\in\cK(t)$, every matched worker $i$ and every matched firm $j$,
\begin{align*}
u^\vee_i(t')&=\max\{u_i(t'),\hat u_i(t')\}, & v^\vee_j(t')&=\min\{v_j(t'),\hat v_j(t')\},\\
u^\wedge_i(t')&=\min\{u_i(t'),\hat u_i(t')\}, & v^\wedge_j(t')&=\max\{v_j(t'),\hat v_j(t')\}.
\end{align*}
Unmatched agents earn zero in all four outcomes. So the two new outcomes have
the payoffs of a join and of a meet at $t$, and it remains to show that they are
Bayesian stable. Fix $t'\in\cK(t)$, and call $(\mu,p)$ and $(\hat\mu,\hat p)$ the
\emph{original} outcomes.

\emph{Individual rationality.} Each agent's payoff at $t'$ in a new outcome
equals his payoff at $t'$ in one of the original outcomes, which is nonnegative.

\emph{No block with an unmatched worker.} Let $i$ be unmatched and let $j$ be a
firm. Worker $i$ is unmatched in all four outcomes, so the willingness set of an
offer $q$ from $j$ to $i$ is the same in all of them. In a new outcome, firm $j$
has the partner and the wage that it has in one of the original outcomes: the
one with which the new outcome coincides on $j$'s component if $j$ is matched,
and either one if $j$ is unmatched. Its re-evaluated payoff on any event is
therefore the same in these two outcomes. If
$(i,j)$ blocked the new outcome at $t'$ with the offer $q$, it would block that
original outcome at $t'$ with the same offer.

\emph{No block with a matched worker: the join.} Suppose that a matched worker
$i$ and a firm $j\ne\mu^\vee(i)$ block $(\mu^\vee,p^\vee,t',\Pi)$ with the offer
$q$. By Lemma~\ref{lem:noscreen} the willingness set is $T$, so firm $j$'s
re-evaluated payoff is its payoff by Lemma~\ref{lem:reeval}(ii), and
\[
a_{ij}(t')+q>u^\vee_i(t')
\qquad\text{and}\qquad
\E\bigl[b_{ij}\mid\Pi_j(t')\bigr]-q>v^\vee_j(t').
\]
Let $O$ be an original outcome in which firm $j$'s payoff at $t'$ equals
$v^\vee_j(t')$; if $j$ is matched, take the outcome with which the join coincides
on $j$'s component. Worker $i$ is not matched with $j$ in $O$: otherwise $i$
would belong to $j$'s component and would be matched with $j$ under $\mu^\vee$.
Since $u^\vee_i(t')$ is at least worker $i$'s payoff at $t'$ in $O$, the worker
strictly gains from $q$ at $t'$ in $O$. By Lemma~\ref{lem:noscreen} the willingness
set in $O$ is then again $T$, so firm $j$'s re-evaluated payoff in $O$ is
$v^\vee_j(t')$ and firm $j$ strictly gains as well. Thus $(i,j)$
blocks $O$ at $t'$, a contradiction.

\emph{No block with a matched worker: the meet.} Suppose that a matched worker
$i$ and a firm $j\ne\mu^\wedge(i)$ block $(\mu^\wedge,p^\wedge,t',\Pi)$ with the
offer $q$. As before, $a_{ij}(t')+q>u^\wedge_i(t')$ and
$\E[b_{ij}\mid\Pi_j(t')]-q>v^\wedge_j(t')$. Let $O$ be the original outcome with
which the meet coincides on $i$'s component, so that $u^\wedge_i(t')$ is worker
$i$'s payoff at $t'$ in $O$ and worker $i$ is not matched with $j$ in $O$. Firm
$j$'s payoff at $t'$ in $O$ is at most $v^\wedge_j(t')$: this is the displayed
identity if $j$ is matched, and both are zero if $j$ is unmatched. The willingness
set of $q$ in $O$ is again $T$ by Lemma~\ref{lem:noscreen}, so this payoff is
also firm $j$'s re-evaluated payoff in $O$. Hence both
strictly gain from $q$ at $t'$ in $O$, and $(i,j)$ blocks $O$ at $t'$, a
contradiction.
\end{proof}

\begin{proof}[Proof of Theorem~\ref{thm:lattice}]
The first claim is Proposition~\ref{prop:lattice}. For the second, fix a
matching. Individual rationality on $\cK(t)$ imposes closed affine constraints
on the wage profile. To see why the no-blocking constraints are closed, fix a
worker--firm pair and a type profile. There are only finitely many possible
willingness sets $D$. For fixed $D$, the conditional expectations are fixed, and
the blocking conditions are strict inequalities in continuous piecewise-affine
functions of the state wage profile and the offered wage. If necessary, lower
the offered wage slightly so that every type outside $D$ is strictly unwillingness;
all strict gains are preserved. A blocking offer therefore remains a blocking
offer after every sufficiently small change in the state wages and the offer,
with the same willingness set. Thus the wage profiles admitting a block form an
open set, and those with no block form a closed set. The wage profile is also
bounded: for a matched pair, worker individual rationality bounds its wage from
below and firm individual rationality bounds it from above; wages attached to
unmatched agents can be normalized to zero. There are finitely many matchings,
so each worker's payoff at $t$ attains a maximum over $S^{B}(t,\Pi)$.
Take, for each worker, an outcome that attains his maximum, and take the join
of these finitely many outcomes. By Proposition~\ref{prop:lattice} it belongs to
$S^{B}(t,\Pi)$, and it gives every worker his maximal payoff. By
Lemma~\ref{lem:common}, an outcome that every worker weakly prefers to another
is weakly worse for every firm. So this outcome is worker-optimal. The
firm-optimal outcome is obtained in the same way from meets.
\end{proof}

\subsection{Proofs for Section~\ref{sec:com}}

\begin{proof}[Proof of Lemma~\ref{lem:noincumbent}]
Write $C:=\Pi_j(t')$. By posterior independence, $\beta(\cdot\mid C)$ is a
product distribution across workers. The event $E=E_i\times T_{-i}$ restricts
only worker $i$'s type. Consequently, conditional on $C\cap E$, the marginal
distribution of every other worker $i'$'s type is unchanged from its marginal
distribution conditional on $C$. Since $b_{i'j}$ depends on the type profile
only through worker $i'$'s type, the two expectations are equal. For the second
claim, the willingness set $D_{ij}(q)$ restricts only worker $i$'s type. If firm $j$
is matched, its worker $\mu(j)$ is not worker $i$, and the first claim applies
with $i'=\mu(j)$. If firm $j$ is unmatched, both payoffs are zero.
\end{proof}

\begin{proof}[Proof of Lemma~\ref{lem:selection}]
If $E\subseteq U$ there is nothing to prove. Otherwise, take $s\in E\cap U$ and
$s'\in E\setminus U$. Then
$a_{ij}(s)-a_{i\ell}(s)\ge c>a_{ij}(s')-a_{i\ell}(s')$, so $b_{ij}(s)\ge b_{ij}(s')$
by Assumption~\ref{ass:com}. Hence
$\E[b_{ij}\mid E\cap U]\ge\E[b_{ij}\mid E\setminus U]$. Since $\E[b_{ij}\mid E]$ is
a weighted average of these two conditional expectations, the claim follows.
\end{proof}

The extension proofs make three substitutions in the separability argument:
Lemma~\ref{lem:oppcom} replaces pairwise opposition, Lemma~\ref{lem:constant}
replaces common differences, and Lemma~\ref{lem:noincumbent} controls
re-evaluated incumbent payoffs when an offer screens the moving worker's types.

In the proofs of Lemma~\ref{lem:oppcom}, Lemma~\ref{lem:constant} and
Theorem~\ref{thm:latticecom}, Assumptions~\ref{ass:com} and~\ref{ass:ind} hold,
and $(\mu,p)$ and $(\hat\mu,\hat p)$ are outcomes in $S^{B}(t,\Pi)$ with payoffs
$(u,v)$ and $(\hat u,\hat v)$. By Lemma~\ref{lem:noincumbent}, in every state
considered in these proofs and for every offer, the re-evaluated payoff of the
firm that makes the offer equals its payoff. A pair therefore blocks a state at
a type profile with an offer if and only if the worker strictly gains there and
the firm's expected value of the worker conditional on its cell and the willingness
set, net of the wage, exceeds its payoff.

\begin{proof}[Proof of Lemma~\ref{lem:oppcom}]
Suppose that $\hat u_i(t)+\hat v_j(t)<S_{ij}(t)$. Consider the state
$(\hat\mu,\hat p,t,\Pi)$ and the offer
$q_\varepsilon:=\hat u_i(t)-a_{ij}(t)+\varepsilon$ from $j$ to $i$, where
$\varepsilon>0$. Since $\hat u_i(s)=a_{i\ell}(s)+\hat p_i$ for every $s\in T$, a
type profile $s$ belongs to $\hat D_{ij}(q_\varepsilon)$ if and only if
\[
a_{ij}(s)-a_{i\ell}(s)>a_{ij}(t)-a_{i\ell}(t)-\varepsilon .
\]
Because $T$ is finite, for every sufficiently small $\varepsilon$ the willingness set
is
\[
\hat D_{ij}(q_\varepsilon)=U:=\bigl\{s\in T:\ a_{ij}(s)-a_{i\ell}(s)\ge a_{ij}(t)-a_{i\ell}(t)\bigr\},
\]
which contains $t$. By Lemma~\ref{lem:selection} with $E=\Pi_j(t)$,
\[
\E\bigl[b_{ij}\mid\Pi_j(t)\cap U\bigr]-q_\varepsilon
\ \ge\ \E\bigl[b_{ij}\mid\Pi_j(t)\bigr]+a_{ij}(t)-\hat u_i(t)-\varepsilon
\ =\ S_{ij}(t)-\hat u_i(t)-\varepsilon .
\]
For $\varepsilon$ small enough the right-hand side exceeds $\hat v_j(t)$, which
equals firm $j$'s re-evaluated payoff $\hat v_j(t\mid U)$ by
Lemma~\ref{lem:noincumbent}. The
worker strictly gains at $t$, since $a_{ij}(t)+q_\varepsilon=\hat u_i(t)+\varepsilon$.
Thus $(i,j)$ blocks $(\hat\mu,\hat p,t,\Pi)$, a contradiction.
\end{proof}

\begin{proof}[Proof of Theorem~\ref{thm:firmcom}]
The proof of Theorem~\ref{thm:firm} uses Assumption~\ref{ass:sep} only through
Lemma~\ref{lem:opposition}. Replacing that lemma by Lemma~\ref{lem:oppcom}, the
proof applies without change.
\end{proof}

\begin{proof}[Proof of Proposition~\ref{prop:necessitycom}]
Since the prior has full support, $b_A(H)<\E[b_A]$. Together with
$\delta_H>\delta_L$, there is a wage $p_A$ with
\[
\max\bigl\{b_A(H),\ \E[b_A]-(\delta_H-\delta_L)\bigr\}<p_A<\E[b_A].
\]
Then $\E[b_A]+\delta_L$, $b_A(H)+\delta_H$ and $p_A+\delta_L$ are all smaller than
$p_A+\delta_H$, so there is a wage $p_B$ with
\[
\max\bigl\{\E[b_A]+\delta_L,\ b_A(H)+\delta_H,\ p_A+\delta_L\bigr\}<p_B<p_A+\delta_H .
\]
These choices give
\[
\delta_L<p_B-p_A<\delta_H,\qquad p_A<\E[b_A]<p_B-\delta_L,\qquad b_A(H)<p_B-\delta_H .
\]
Next choose values of firm $B$ with
\[
p_B<\E[b_B]<p_A+\delta_H\qquad\text{and}\qquad b_B(L)<p_A+\delta_L ,
\]
which is possible because $p_B<p_A+\delta_H$ and the prior has full support: fix
$b_B(L)$ first and then choose $b_B(H)$ to obtain the required expectation.
Finally let $a_B(s):=c$ and $a_A(s):=c+\delta_s$ for both types $s$, where $c$ is
large enough that $a_A(s)+p_A>0$ and $a_B(s)+p_B>0$.

The wages and the values now satisfy every inequality displayed in the proof
of Proposition~\ref{prop:necessity}. That proof shows that the outcome in which
the worker is matched with $A$ at the wage $p_A$ and the outcome in which he is
matched with $B$ at the wage $p_B$ are both in the stratum at each $t$. Firm $A$
earns $\E[b_A]-p_A>0$ in the first and is unmatched in the second. The
inequalities imposed on firm $B$'s values are strict, so they hold on an open
set.
\end{proof}

\begin{proof}[Proof of Proposition~\ref{prop:necessityind}]
All firms are uninformed, so each state must be individually rational and not
blocked at every type profile. In (a) the workers earn $10$, $15$ and $12$ at
every type profile, and firms $A$, $B$ and $C$ earn $2$, $26-5=21$ and
$20-2=18$. In (b) worker $1$ earns $13$, worker $2$ earns $10$ at type $L$ and
$20$ at type $H$, worker $3$ earns $10$, and firms $B$, $C$ and $D$ earn
$20-3=17$, $22$ and $1$. Both states are individually rational.

Consider a pair $(i,j)$ with $a_{ij}=b_{ij}=-100$. Worker $i$ earns at least
$10$ in both outcomes, so he accepts an offer $q$ only if $q>110$, and firm $j$
would then earn less than $-210$. Firms' values of the workers they employ are
nonnegative and wages are at most $5$, so a firm's re-evaluated payoff is at
least $-5$ on every event. No such pair blocks. Six pairs remain. Each is
matched in one of the two outcomes and can block only the other.

\emph{Outcome (a).} Both types of worker $1$ accept an offer $q$ from $B$ if and
only if $q>0$, and $B$ would earn $20-q<20$, less than its payoff $21$. Both
types of worker $3$ accept an offer $q$ from $D$ if and only if $q>2$, and $D$
would earn $1-q<0$. Worker $2$ of type $H$ accepts an offer $q$ from $C$ if and
only if $20+q>15$, i.e., $q>-5$, and type $L$ if and only if $q>5$. If
$-5<q\le5$, the willingness set is the event that worker $2$ has type $H$. Firm $C$
would earn $22-q<27$. Its re-evaluated payoff on this event is $30-2=28$, so
$(2,C)$ does not block. If $q>5$, both types accept, and $C$ would earn
$22-q<17$, less than its payoff $18$.

\emph{Outcome (b).} Both types of worker $1$ accept an offer $q$ from $A$ if and
only if $q>3$, and $A$ would earn $2-q<0$. Both types of worker $3$ accept an
offer $q$ from $C$ if and only if $q>0$, and $C$ would earn $20-q<20$, less than
its payoff $22$. Worker $2$ of type $L$ accepts an offer $q$ from $B$ if and
only if $q>0$, and type $H$ if and only if $10+q>20$, i.e., $q>10$. If
$0<q\le10$, the willingness set is the event that worker $2$ has type $L$. Firm $B$
would earn $26-q<26$. Its re-evaluated payoff on this event is $30-3=27$, so
$(2,B)$ does not block. If $q>10$, both types accept, and $B$ would earn
$26-q<16$, less than its payoff $17$.
\end{proof}

\begin{proof}[Proof of Theorem~\ref{thm:genericcom}]
The argument of Section~\ref{sec:workers} uses Assumption~\ref{ass:sep} only
through pairwise opposition, so it applies with Lemma~\ref{lem:oppcom} and yields
the same hyperplanes. Assumption~\ref{ass:ind} restricts $\beta$ and $\Pi$ and
not firm values. Each hyperplane is defined by an equation in which the
interim value $\E[b_{ij}\mid\Pi_j(t)]$ of some pair $(i,j)$ appears exactly once.
Adding a constant to $b_{ij}(s)$ for every $s\in T_i$ changes that equation and
preserves Assumption~\ref{ass:com}, which only compares the values of one worker
at one firm across his types. Hence no hyperplane contains $\calB$.
\end{proof}

\begin{proof}[Proof of Lemma~\ref{lem:constant}]
If $C=\{i,j\}$ is matched under both matchings, both differences equal
$\hat p_i-p_i$ at every type profile. Suppose that $C$ is a cycle
$i_1,j_1,\dots,i_k,j_k$ with $k\ge2$, $\mu(i_r)=j_r$ and $\hat\mu(i_r)=j_{r+1}$,
indices being taken modulo $k$. For $t'\in\cK(t)$ write
$\Delta u_i(t'):=\hat u_i(t')-u_i(t')$ and $\Delta v_j(t'):=v_j(t')-\hat v_j(t')$.

The logic of the remaining steps is as follows. A strict change in $d$ orders
the relative eagerness of the two workers employed by a firm in the two outcomes
in opposite directions. Comonotonicity then orders the firm's values of these
two workers in opposite directions. The resulting inequalities contradict the
expression for the change in the firm's loss.

\emph{Step 1: at each $t'\in\cK(t)$ the differences are equal on $C$.} Since
$\cK(t')=\cK(t)$, both outcomes belong to $S^{B}(t',\Pi)$.
Lemma~\ref{lem:oppcom} at $t'$, applied to $(i_r,j_r)$, gives
$\Delta u_{i_r}(t')\ge\Delta v_{j_r}(t')$; applied to $(i_r,j_{r+1})$ with the
roles of the outcomes exchanged, it gives
$\Delta v_{j_{r+1}}(t')\ge\Delta u_{i_r}(t')$. Going around the cycle,
\[
\Delta v_{j_1}(t')\le\Delta u_{i_1}(t')\le\Delta v_{j_2}(t')\le\dots\le\Delta u_{i_k}(t')\le\Delta v_{j_1}(t').
\]
Let $d(t')$ denote the common value.

\emph{Step 2: $d$ is constant on each cell of each firm of $C$.} A firm's payoff
at $t'$ depends on $t'$ only through its cell, so $d(s)=\Delta v_j(s)=\Delta v_j(t')$
for every firm $j\in C$ and every $s\in\Pi_j(t')$.

\emph{Step 3: $d$ is constant on $\cK(t)$.} Suppose that $d(t')>d(t'')$ for some
$t',t''\in\cK(t)$. Fix a firm $j=j_r$ of $C$, and let $i:=i_r=\mu(j)$,
$i':=i_{r-1}=\hat\mu(j)$, $\hat\jmath:=\hat\mu(i)$ and $j^\circ:=\mu(i')$. Take any
$s\in\Pi_j(t')$ and $s''\in\Pi_j(t'')$. By Step 2, $d(s)>d(s'')$.

Since $\Delta u_i(s)=a_{i\hat\jmath}(s)-a_{ij}(s)+\hat p_i-p_i$, the inequality
$\Delta u_i(s)>\Delta u_i(s'')$ reads
$a_{ij}(s'')-a_{i\hat\jmath}(s'')>a_{ij}(s)-a_{i\hat\jmath}(s)$, and
Assumption~\ref{ass:com} gives $b_{ij}(s'')\ge b_{ij}(s)$. As this holds for all
such $s$ and $s''$, it follows that
$\E[b_{ij}\mid\Pi_j(t'')]\ge\E[b_{ij}\mid\Pi_j(t')]$. Since
$\Delta u_{i'}(s)=a_{i'j}(s)-a_{i'j^\circ}(s)+\hat p_{i'}-p_{i'}$, the inequality
$\Delta u_{i'}(s)>\Delta u_{i'}(s'')$ reads
$a_{i'j}(s)-a_{i'j^\circ}(s)>a_{i'j}(s'')-a_{i'j^\circ}(s'')$, and
Assumption~\ref{ass:com} gives $b_{i'j}(s)\ge b_{i'j}(s'')$. Hence
$\E[b_{i'j}\mid\Pi_j(t')]\ge\E[b_{i'j}\mid\Pi_j(t'')]$. Firm $j$ employs $i$ under
$\mu$ and $i'$ under $\hat\mu$, so the wages cancel in
\[
d(t')-d(t'')=\Bigl(\E[b_{ij}\mid\Pi_j(t')]-\E[b_{ij}\mid\Pi_j(t'')]\Bigr)
-\Bigl(\E[b_{i'j}\mid\Pi_j(t')]-\E[b_{i'j}\mid\Pi_j(t'')]\Bigr)\ \le\ 0,
\]
a contradiction.
\end{proof}

\begin{proof}[Proof of Theorem~\ref{thm:latticecom}]
(i) Let $(\mu,p)$ and $(\hat\mu,\hat p)$ be the two outcomes and let $d_C$ be as
in Lemma~\ref{lem:constant}. Define $(\mu^\vee,p^\vee)$ as follows: on each
component $C$ with $d_C>0$ it coincides with $(\hat\mu,\hat p)$, on every other
component it coincides with $(\mu,p)$, and the agents who are unmatched under
$\mu$ are unmatched. Define $(\mu^\wedge,p^\wedge)$ in the same way with $d_C<0$
in place of $d_C>0$. Both are well defined, because each matching matches the
agents of a component among themselves. By Lemma~\ref{lem:constant}, at every
$t'\in\cK(t)$ the first outcome gives each matched worker the larger and each
matched firm the smaller of their two payoffs, and the second gives the
reverse. In particular they have the payoffs of a join and of a meet at $t$. It
remains to show that both are Bayesian stable. Fix $t'\in\cK(t)$, and say that a
new outcome \emph{copies} an original outcome on a component if it coincides
with it there.

\emph{Individual rationality.} Each agent's payoff at $t'$ in a new outcome
equals his payoff at $t'$ in one of the original outcomes.

\emph{No block with an unmatched worker.} An unmatched worker $i$ is unmatched
in all four outcomes, so the willingness set of an offer from a firm $j$ is the same
in all of them, and in a new outcome firm $j$ has the partner and the wage that
it has in one of the original outcomes. A block of the new outcome would
therefore block that original outcome.

\emph{No block with a matched worker: the meet.} Suppose that a matched worker
$i$ and a firm $j\ne\mu^\wedge(i)$ block the meet at $t'$ with the offer $q$. Let
$O$ be the outcome that the meet copies on $i$'s component. Worker $i$ has the
same partner and the same wage in $O$ as in the meet, so the willingness set of $q$
is the same in both, the worker strictly gains at $t'$ in $O$, and he is not
matched with $j$ in $O$. Firm $j$'s payoff at $t'$ in $O$ is at most its payoff in
the meet. So $(i,j)$ blocks $O$ at $t'$, a contradiction.

\emph{No block with a matched worker: the join.} Suppose that a matched worker
$i$ and a firm $j\ne\mu^\vee(i)$ block the join at $t'$ with the offer $q$. Let
$O_i$ be the outcome that the join copies on $i$'s component, with willingness set
$D^{O_i}_{ij}(q)$, and let $O_j$ be an original outcome in which firm $j$'s payoff
at $t'$ equals its payoff in the join; if $j$ is matched, take the outcome that
the join copies on $j$'s component. Then $t'\in D^{O_i}_{ij}(q)$, and firm $j$'s
expected value of worker $i$ conditional on $\Pi_j(t')\cap D^{O_i}_{ij}(q)$, net of
$q$, exceeds its payoff at $t'$ in $O_j$. Worker $i$ is not matched with $j$ in
$O_j$: otherwise $i$ would belong to $j$'s component and would be matched with
$j$ in the join. If $O_j=O_i$, the pair $(i,j)$ blocks $O_i$ at $t'$, a
contradiction. Suppose that $O_j\neq O_i$. The join copies on $i$'s component the
outcome that its workers weakly prefer, so by Lemma~\ref{lem:constant} there is
$d\ge0$ such that worker $i$'s payoff in $O_j$ equals his payoff in $O_i$ minus
$d$ at every type profile in $\cK(t)$. Consider the offer $q-d$ in $O_j$. A type
profile $s\in\cK(t)$ belongs to its willingness set if and only if
$s\in D^{O_i}_{ij}(q)$. Since $\Pi_j(t')\subseteq\cK(t)$, the worker strictly gains
at $t'$, firm $j$ conditions on the same event, and it pays $d$ less. So $(i,j)$
blocks $O_j$ at $t'$, a contradiction.

(ii) The first claim follows from (i). For the second, the closedness and
boundedness argument in the proof of Theorem~\ref{thm:lattice} applies verbatim
here, so each worker's payoff at $t$ attains a maximum over $S^{B}(t,\Pi)$. The join of
finitely many outcomes, one attaining each worker's maximum, is worker-optimal,
since by Lemma~\ref{lem:constant} an outcome that every worker weakly prefers to
another is weakly worse for every firm. The firm-optimal outcome is obtained in
the same way from meets.
\end{proof}

\subsection{Proofs for Section~\ref{sec:ntu}}\label{app:ntu}

\begin{proof}[Proof of Proposition~\ref{prop:relations}]
(i) Under Assumption~\ref{ass:sep}, $a_{ij}(s)-a_{i\ell}(s)=h_{ij}-h_{i\ell}$ for
every $s$. The premise of Assumption~\ref{ass:com} never holds, and the sign of
the difference does not depend on $s$. (ii) The sign of
$a_{ij}(s)-a_{i\ell}(s)=g_i(s)(h_{ij}-h_{i\ell})$ is the sign of $h_{ij}-h_{i\ell}$.
(iii) If $a_{ij}(s)>a_{i\ell}(s)$ and $a_{ij}(s')\le a_{i\ell}(s')$, then
$a_{ij}(s)-a_{i\ell}(s)>0\ge a_{ij}(s')-a_{i\ell}(s')$, and Assumption~\ref{ass:com}
gives $b_{ij}(s)\ge b_{ij}(s')$. (iv) Under type-independent rankings the premise
of ordinal \com{} never holds.

For the last claim, consider one worker with types $L$ and $H$ and two firms
$A$ and $B$, and write $a(s)=(a_A(s),a_B(s))$. If $a(L)=(2,1)$, $a(H)=(4,2)$ and
$b_A(L)>b_A(H)$, values are multiplicatively separable, but they are neither
additively separable nor comonotone. If $a(L)=(1,2)$, $a(H)=(2,3)$, values are
additively separable and not multiplicatively separable. If $a(L)=(1,2)$,
$a(H)=(2,1)$ and firms' values do not depend on the type, values are comonotone,
and rankings depend on the type. These three examples rule out every
implication not listed.
\end{proof}

The rest of this appendix proves Proposition~\ref{prop:ntu}. The model is the one without
transfers described in Section~\ref{sec:ntu}. Write
$B_{ij}(t'):=\E[b_{ij}\mid\Pi_j(t')]$ for firm $j$'s interim value of worker $i$,
and $a_{ii}\equiv0$ and $B_{jj}\equiv0$ for unmatched agents. Since
$\cK(t')=\cK(t)$ for every $t'\in\cK(t)$, the sets $S^{B}(t',\Pi)$ and
$S^{B}(t,\Pi)$ coincide for every $t'\in\cK(t)$.

Fix $t$ and $\Pi$, two matchings $\mu,\hat\mu\in S^{B}(t,\Pi)$, and $t'\in\cK(t)$.
Strict preferences are maintained throughout. By individual rationality and
strictness, at every $s\in\cK(t)$,
\begin{equation}\label{eq:ntuIR}
a_{i\mu(i)}(s)>0\ \text{ if }\mu(i)\in J,
\qquad
B_{\mu(j)j}(s)>0\ \text{ if }\mu(j)\in I,
\end{equation}
and the same holds for $\hat\mu$.

\begin{lemma}[Selection without transfers]\label{lem:ntusel}
Suppose ordinal \com{} holds. Let $i\in I$, $j\ne\ell$ in $J$, and
$D:=\{s\in T:a_{ij}(s)>a_{i\ell}(s)\}$. For every event $E$ with
$E\cap D\ne\emptyset$, $\E[b_{ij}\mid E\cap D]\ge\E[b_{ij}\mid E]$.
\end{lemma}
\begin{proof}
If $E\subseteq D$ there is nothing to prove. Otherwise, for $s\in E\cap D$ and
$s'\in E\setminus D$, ordinal \com{} gives $b_{ij}(s)\ge b_{ij}(s')$. Hence
$\E[b_{ij}\mid E\cap D]\ge\E[b_{ij}\mid E\setminus D]$, and $\E[b_{ij}\mid E]$ is a
weighted average of the two.
\end{proof}

\begin{lemma}[Rehiring without transfers]\label{lem:nturehire}
If $\mu(i)=j\in J$ and $i$ is unmatched under $\hat\mu$, then
$B_{\hat\mu(j)j}(t')>B_{ij}(t')>0$. In particular, $j$ is matched under $\hat\mu$.
\end{lemma}
\begin{proof}
By \eqref{eq:ntuIR}, $a_{ij}(s)>0$ for every $s\in\cK(t)$, so
$\cK(t)\subseteq D_{\hat\mu,ij}$ and $\Pi_j(t')\cap D_{\hat\mu,ij}=\Pi_j(t')$. The
worker is willingness at $t'$. Since $(i,j)$ does not block $(\hat\mu,t',\Pi)$,
$B_{ij}(t')\le B_{\hat\mu(j)j}(t')$. As $\hat\mu(j)\ne i$, the inequality is strict
by strict preferences. Finally $B_{ij}(t')>0$ by \eqref{eq:ntuIR}, so
$\hat\mu(j)\ne j$.
\end{proof}

\begin{lemma}[Opposition without transfers]\label{lem:ntuopp}
Suppose that either workers have type-independent rankings, or ordinal \com{} and Assumption~\ref{ass:ind} hold. If $\mu(i)=j\in J$, $\hat\mu(i)=\hat\jmath\in J$ with
$\hat\jmath\ne j$, and $a_{ij}(t')>a_{i\hat\jmath}(t')$, then
$B_{\hat\mu(j)j}(t')>B_{ij}(t')>0$. In particular, $j$ is matched under $\hat\mu$.
\end{lemma}
\begin{proof}
Let $D:=D_{\hat\mu,ij}=\{s:a_{ij}(s)>a_{i\hat\jmath}(s)\}$, which contains $t'$, and
let $i':=\hat\mu(j)$; note $i'\neq i$. Since $(i,j)$ does not block
$(\hat\mu,t',\Pi)$,
\begin{equation}\label{eq:ntunb}
\E\bigl[b_{ij}\mid\Pi_j(t')\cap D\bigr]\le\E\bigl[b_{i'j}\mid\Pi_j(t')\cap D\bigr].
\end{equation}
Under type-independent rankings, $D=T$, and \eqref{eq:ntunb} reads
$B_{ij}(t')\le B_{i'j}(t')$. Under ordinal \com{} and Assumption~\ref{ass:ind}, the left-hand
side of \eqref{eq:ntunb} is at least $B_{ij}(t')$ by Lemma~\ref{lem:ntusel}. The right-hand side
is $0$ if $i'=j$. If $i'\in I$, it equals $B_{i'j}(t')$: the event $D$ restricts only
worker $i$'s type, so posterior independence across workers leaves the
conditional marginal distribution of worker $i'$'s type unchanged
(Lemma~\ref{lem:noincumbent}). In both cases
$B_{ij}(t')\le B_{i'j}(t')$. The inequality is strict by strict preferences, and
$B_{ij}(t')>0$ by \eqref{eq:ntuIR}, so $i'\ne j$.
\end{proof}

Say that a matched firm $j$ \emph{prefers $\mu$ at $t'$} if
$B_{\mu(j)j}(t')>B_{\hat\mu(j)j}(t')$, and that a worker $i$ \emph{prefers $\mu$ at
$t'$} if $a_{i\mu(i)}(t')>a_{i\hat\mu(i)}(t')$; similarly for $\hat\mu$.

\begin{corollary}\label{cor:ntuopp}
Under the hypotheses of Lemma~\ref{lem:ntuopp}: if $\mu(j)=i\in I$ and firm $j$ prefers
$\mu$ at $t'$, then worker $i$ is matched under $\hat\mu$ and prefers $\hat\mu$ at
$t'$.
\end{corollary}
\begin{proof}
The inequality $B_{ij}(t')>B_{\hat\mu(j)j}(t')$ implies $\hat\mu(j)\neq i$. If $i$ were
unmatched under $\hat\mu$, Lemma~\ref{lem:nturehire} would give
$B_{\hat\mu(j)j}(t')>B_{ij}(t')$. If $i$ were matched under $\hat\mu$ and preferred
$\mu$ at $t'$, Lemma~\ref{lem:ntuopp} would give the same inequality. By
strict preferences the worker is not indifferent.
\end{proof}

Lemmas~\ref{lem:nturehire} and~\ref{lem:ntuopp} and Corollary~\ref{cor:ntuopp} also hold with the roles of
$\mu$ and $\hat\mu$ exchanged.

\begin{proof}[Proof of Proposition~\ref{prop:ntu}(i)]
Let $\mu,\hat\mu\in S^{B}(t,\Pi)$ and apply the lemmas at $t'=t$.

\emph{Workers.} Suppose that a worker $i_0$ is matched under $\mu$ and unmatched
under $\hat\mu$, and let $j_1:=\mu(i_0)$. I construct distinct agents
$i_0,j_1,i_1,j_2,i_2,\dots$ with $\mu(i_{r-1})=j_r$ and $\hat\mu(j_r)=i_r$ such
that, for every $r\ge1$, firm $j_r$ prefers $\hat\mu$ at $t$ and worker $i_r$ is
matched under $\mu$ and prefers $\mu$ at $t$. By Lemma~\ref{lem:nturehire},
$i_1:=\hat\mu(j_1)\in I$ and $j_1$ prefers $\hat\mu$. Suppose that $j_r$ prefers
$\hat\mu$ and $i_r=\hat\mu(j_r)\in I$. By Corollary~\ref{cor:ntuopp} with the roles
exchanged, $i_r$ is matched under $\mu$ and prefers $\mu$. Let
$j_{r+1}:=\mu(i_r)\ne j_r$. By Lemma~\ref{lem:ntuopp} applied to $i_r$,
$i_{r+1}:=\hat\mu(j_{r+1})\in I$ and $j_{r+1}$ prefers $\hat\mu$. The agents
alternate along edges of $\mu$ and $\hat\mu$, and $i_0$ has no partner under
$\hat\mu$, so they are distinct. This contradicts finiteness.

\emph{Firms.} Suppose that a firm $j_0$ is matched under $\mu$ and unmatched
under $\hat\mu$, and let $i_1:=\mu(j_0)$. By \eqref{eq:ntuIR},
$B_{i_1j_0}(t)>0=B_{\hat\mu(j_0)j_0}(t)$, so $j_0$ prefers $\mu$. I construct
distinct agents $j_0,i_1,j_1,i_2,j_2,\dots$ with $\mu(j_{r-1})=i_r$ and
$\hat\mu(i_r)=j_r$ such that every $j_{r-1}$ prefers $\mu$ and every $i_r$ is
matched under $\hat\mu$ and prefers $\hat\mu$. Suppose that $j_{r-1}$ prefers $\mu$
and $i_r=\mu(j_{r-1})\in I$. By Corollary~\ref{cor:ntuopp}, $i_r$ is matched under
$\hat\mu$ and prefers $\hat\mu$. Let $j_r:=\hat\mu(i_r)\ne j_{r-1}$. By
Lemma~\ref{lem:ntuopp} with the roles exchanged, applied to $i_r$,
$i_{r+1}:=\mu(j_r)\in I$ and $j_r$ prefers $\mu$. Again the agents are distinct,
because $j_0$ has no partner under $\hat\mu$, which contradicts finiteness.

Exchanging $\mu$ and $\hat\mu$ completes the proof.
\end{proof}

Recall from Section~\ref{sec:ntu} that $\lambda=\mu\vee_I\hat\mu$ assigns to
each worker his more preferred partner at $t$ and to each firm the partner with
the lower value of $B_{\cdot j}(t)$, and that $\nu=\mu\wedge_I\hat\mu$ assigns to
each worker his less preferred partner and to each firm the partner with the
higher value.

\begin{proof}[Proof of Proposition~\ref{prop:ntu}(ii)]
By part~(i), $\mu$ and $\hat\mu$ have the same unmatched agents. Let
$I_\mu$ be the set of workers who have different partners under the two
matchings and prefer $\mu$, and define $I_{\hat\mu}$ analogously. By
type-independent rankings these sets do not depend on the type profile, and by
strict preferences every worker with different partners belongs to one of them.
For $t'\in\cK(t)$, let $J_{\hat\mu}(t')$ be the set of firms with different
partners that prefer $\hat\mu$ at $t'$, and define $J_\mu(t')$ analogously; by
strict preferences every firm with different partners belongs to exactly one of
them.

\emph{Step 1: decomposition.} Fix $t'\in\cK(t)$. By Lemma~\ref{lem:ntuopp},
$\mu(I_\mu)\subseteq J_{\hat\mu}(t')$. By Corollary~\ref{cor:ntuopp} with the roles
exchanged, $\hat\mu(J_{\hat\mu}(t'))\subseteq I_\mu$. Since $\mu$ and $\hat\mu$ are
one-to-one, $|I_\mu|\le|J_{\hat\mu}(t')|\le|I_\mu|$, so
\[
J_{\hat\mu}(t')=\mu(I_\mu)=\hat\mu(I_\mu)=:J_{\hat\mu}
\qquad\text{for every } t'\in\cK(t).
\]
Symmetrically, $J_\mu(t')=\mu(I_{\hat\mu})=\hat\mu(I_{\hat\mu})=:J_\mu$ for every
$t'\in\cK(t)$. Thus a firm's ranking of its two partners is the same at every type
profile in $\cK(t)$.

\emph{Step 2: $\lambda$ and $\nu$ are matchings.} The function $\lambda$ coincides
with $\mu$ on $I_\mu\cup J_{\hat\mu}$ and with $\hat\mu$ on $I_{\hat\mu}\cup J_\mu$,
and with both on all other agents. By Step 1, $\mu$ matches $I_\mu$ with
$J_{\hat\mu}$ and $\hat\mu$ matches $I_{\hat\mu}$ with $J_\mu$, so $\lambda$ is a
matching. Likewise $\nu$ coincides with $\hat\mu$ on $I_\mu\cup J_{\hat\mu}$ and
with $\mu$ on $I_{\hat\mu}\cup J_\mu$, and is a matching. Moreover, for every
$s\in T$ and every $t'\in\cK(t)$,
\begin{align*}
a_{i\lambda(i)}(s)&=\max\{a_{i\mu(i)}(s),a_{i\hat\mu(i)}(s)\}, &
B_{\lambda(j)j}(t')&=\min\{B_{\mu(j)j}(t'),B_{\hat\mu(j)j}(t')\},\\
a_{i\nu(i)}(s)&=\min\{a_{i\mu(i)}(s),a_{i\hat\mu(i)}(s)\}, &
B_{\nu(j)j}(t')&=\max\{B_{\mu(j)j}(t'),B_{\hat\mu(j)j}(t')\}.
\end{align*}

\emph{Step 3: individual rationality.} Fix $t'\in\cK(t)$. Every agent's partner
under $\lambda$ or $\nu$ is his partner under $\mu$ or under $\hat\mu$, and both
are individually rational at $t'$.

\emph{Step 4: no blocking pair with an unmatched worker.} Let $\kappa\in\{\lambda,\nu\}$
and suppose that $(i,j)$ blocks $(\kappa,t',\Pi)$, where $i$ is unmatched under
$\kappa$. Then $i$ is unmatched under $\mu$ and $\hat\mu$, so
$D_{\kappa,ij}=D_{\mu,ij}=D_{\hat\mu,ij}$. Let $O\in\{\mu,\hat\mu\}$ be such that
$O(j)=\kappa(j)$. Then $(i,j)$ blocks $(O,t',\Pi)$, a contradiction.

\emph{Step 5: no blocking pair with a matched worker.} Let $\kappa\in\{\lambda,\nu\}$
and suppose that $(i,j)$ blocks $(\kappa,t',\Pi)$, where $\kappa(i)\in J$ and
$\kappa(i)\ne j$. Since $a_{ij}(t')>a_{i\kappa(i)}(t')$, type-independent rankings give
$D_{\kappa,ij}=T$, and the blocking condition reads
\begin{equation}\label{eq:ntublk}
a_{ij}(t')>a_{i\kappa(i)}(t')
\qquad\text{and}\qquad
B_{ij}(t')>B_{\kappa(j)j}(t').
\end{equation}

If $\kappa=\lambda$, let $O\in\{\mu,\hat\mu\}$ be such that $O(j)=\lambda(j)$. By
Step 2, $a_{i\lambda(i)}(t')\ge a_{iO(i)}(t')$, so $a_{ij}(t')>a_{iO(i)}(t')$; in
particular $O(i)\ne j$. By type-independent rankings, $D_{O,ij}=T$, and
$B_{ij}(t')>B_{\lambda(j)j}(t')=B_{O(j)j}(t')$ by \eqref{eq:ntublk}. So $(i,j)$ blocks
$(O,t',\Pi)$, a contradiction.

If $\kappa=\nu$, let $O\in\{\mu,\hat\mu\}$ be such that $O(i)=\nu(i)$, so
$O(i)\neq j$ and $a_{ij}(t')>a_{iO(i)}(t')$. By type-independent rankings,
$D_{O,ij}=T$. By Step 2 and \eqref{eq:ntublk},
$B_{ij}(t')>B_{\nu(j)j}(t')\ge B_{O(j)j}(t')$. So $(i,j)$ blocks $(O,t',\Pi)$, a
contradiction.
\end{proof}

\begin{proof}[Proof of Proposition~\ref{prop:ntu}(iii)]
There are four workers and four firms $A,B,C,D$. Only worker $1$ has private
information, with equally likely types $L$ and $H$, and all firms are
uninformed, so Assumption~\ref{ass:ind} holds. Worker $1$ ranks the firms $B,C,A,D$ at $L$
and $C,B,A,D$ at $H$. Workers $2$, $3$ and $4$ rank them $A,B,C,D$; $C,D,A,B$; and
$D,C,A,B$. Every worker prefers every firm to being unmatched. Firms' values of
worker $1$ are $(8,3,1,1)$ at $L$ and $(8,2,7,1)$ at $H$ for $(A,B,C,D)$, and their
values of workers $2$, $3$ and $4$ are $(5,6,2,2)$, $(1,1,5,6)$ and
$(0.5,0.5,9,4)$. Worker $1$'s types disagree only about $B$ and $C$, and
$b_{1B}(L)=3\ge2=b_{1B}(H)$ and $b_{1C}(H)=7\ge1=b_{1C}(L)$, so ordinal \com{} holds.
Preferences are strict: on every event, a firm's expected value of worker $1$
($8$ at $A$; $3$, $2$ or $2.5$ at $B$; $1$, $7$ or $4$ at $C$; $1$ at $D$)
differs from its values of the other workers, which are distinct and nonzero.

Let $\mu=\{1A,2B,3C,4D\}$ and $\hat\mu=\{1B,2A,3D,4C\}$. Both matchings are
individually rational. The following table lists every pair with a nonempty
willingness set under either matching. ``New value'' is the deviating firm's
expected value of the worker it approaches, conditional on the willingness set;
``incumbent value'' is the value of its current partner on that same event.
\[
\begin{array}{ccccc}
\text{State}&\text{Pair }(i,j)&\text{willingness set}&\text{New value}&\text{Incumbent value}\\ \hline
\mu&(1,B)&T&2.5&6\\
\mu&(1,C)&T&4&5\\
\mu&(2,A)&T&5&8\\
\hat\mu&(1,C)&\{H\}&7&9\\
\hat\mu&(3,C)&T&5&9\\
\hat\mu&(4,D)&T&4&6
\end{array}
\]
Pairs not listed have an empty willingness set and therefore cannot block. For
example, under $\mu$ both types of worker $1$ prefer $C$ to $A$, so firm $C$
expects $4$ from worker $1$, less than the value $5$ of worker $3$. Under
$\hat\mu$ only type $H$ prefers $C$ to $B$, and firm $C$ compares the resulting
value $7$ with the value $9$ of worker $4$. In every row the new value is
smaller than the incumbent value, so neither matching has a blocking pair.
Since all firms are uninformed, information stability also holds. Thus both
matchings belong to $S^{B}(t,\Pi^{0})$ for each $t$.

Workers $1$ and $2$ prefer $\hat\mu$ and workers $3$ and $4$ prefer $\mu$, and
every firm values its partner under $\{1B,2A,3C,4D\}$ less than its other
partner. So $\mu\vee_I\hat\mu=\{1B,2A,3C,4D\}$, which is a matching. It is
blocked at $H$ by worker $1$ and firm $C$: only type $H$ prefers $C$ to $B$,
firm $C$ infers the value $7$, and its partner under the join is worker $3$,
with value $5$.
\end{proof}

\end{document}